\documentclass[a4paper,onecolumn,11pt,accepted=2023-03-27]{quantumarticle}
\pdfoutput=1

\usepackage[utf8]{inputenc}
\usepackage[english]{babel}
\usepackage[T1]{fontenc}

\usepackage[unicode=true]{hyperref}
\usepackage{amsmath}
\usepackage{amsfonts}
\usepackage{amssymb}
\usepackage{physics}
\usepackage[pdftex]{color,graphicx}
\usepackage[table,dvipsnames]{xcolor}
\usepackage{enumitem}

\usepackage[numbers]{natbib}

\newtheorem{Theorem}{Theorem}
\newtheorem{Cor}{Corollary}
\newtheorem{Lemma}{Lemma} 
\newtheorem{Def}{Definition}

\newtheorem{Obs}{Observation}

\newcommand{\ZZ}{\mathbb{Z}}
\newcommand{\CC}{\mathbb{C}}
\newcommand{\cC}{\mathcal{C}}
\newcommand{\oO}{\mathcal{O}}
\newcommand{\Cl}{\text{Cl}}  
\newcommand{\cov}{\text{cov}}  

\newcommand{\Sp}{\text{Sp}}

\newcommand{\veq}{\rotatebox[origin=c]{-90}{$=$}}
\newcommand{\vimpl}{\rotatebox[origin=c]{-90}{$\Rightarrow$}}

\begin{document}

\title{The role of cohomology in quantum computation with magic states}

\author{Robert Raussendorf}
\affiliation{Department of Physics \& Astronomy, University of British Columbia, Vancouver, Canada}
\affiliation{Stewart Blusson Quantum Matter Institute, University of British Columbia, Vancouver, Canada}
\orcid{0000-0003-4983-9213}

\author{Cihan Okay}
\affiliation{Department of Mathematics, Bilkent University, Ankara, Turkey}
\orcid{0000-0001-8097-5227}

\author{Michael Zurel}
\affiliation{Department of Physics \& Astronomy, University of British Columbia, Vancouver, Canada}
\affiliation{Stewart Blusson Quantum Matter Institute, University of British Columbia, Vancouver, Canada}
\orcid{0000-0003-0333-1174}

\author{Polina Feldmann}
\affiliation{Department of Physics \& Astronomy, University of British Columbia, Vancouver, Canada}
\affiliation{Stewart Blusson Quantum Matter Institute, University of British Columbia, Vancouver, Canada}
\orcid{0000-0002-9516-1899}


\maketitle

\begin{abstract}A web of cohomological facts relates quantum error correction, measurement-based quantum computation, symmetry protected topological order and contextuality. Here we extend this web to quantum computation with magic states. In this computational scheme, the negativity of certain quasiprobability functions is an indicator for quantumness. However, when constructing quasiprobability functions to which this statement applies, a marked difference arises  between the cases of even and odd local Hilbert space dimension.  At a technical level, establishing negativity as an indicator of quantumness in quantum computation with magic states relies on two properties of the Wigner function: their covariance with respect to the Clifford group and positive representation of Pauli measurements. In odd dimension, Gross' Wigner function---an adaptation of the original Wigner function to odd-finite-dimensional Hilbert spaces---possesses these properties. 
In even dimension, Gross' Wigner function doesn't exist. Here we discuss the broader class of Wigner functions that, like Gross', are obtained from operator bases. We find that such Clifford-covariant Wigner functions do not exist in any even dimension, and furthermore, Pauli measurements cannot be positively represented by them in any even dimension whenever the number of qudits is  $n\geq 2$. We establish that the obstructions to the existence of such Wigner functions are cohomological.\end{abstract}

\section{Introduction}

Homology and cohomology find widespread uses in physics \cite{Naka}, starting with Gauss' theorem in electromagnetism. They have more recently entered the fields of foundations of quantum mechanics and quantum computation. Regarding the former, Kochen-Specker contextuality~\cite{KS} has recently been given a cohomological underpinning \cite{AbramskyBrandenburger,BMA,Coho}. Regarding the latter, Wigner function negativity---a traditional indicator of nonclassicality in quantum optics~\cite{KenfackZyckowski2004}---has been shown to be equivalent to contextuality in certain cases~\cite{SpekkensEquivalence2008,DelfosseEquivalence2017,JuaniCVEquivalence,EmeriauCVEquivalence}, and both have been linked to the possibility of quantum computational advantage~\cite{Galvao,Cormick,Howard,Entropy,NegWi}.

A short list of cohomological phenomena in the field of quantum computation is as follows: (i) quantum error correction with the Kitaev surface code \cite{Kit,DLKP}, and its measurement-based counterpart with 3D cluster states \cite{RHG06}; (ii) proofs of contextuality of quantum mechanics \cite{KS,AbramskyBrandenburger,BMA,Coho}, in particular parity-based proofs such as Mermin's square and star \cite{Merm}; (iii) the contextuality of measurement-based quantum computation (MBQC) \cite{RB01} in the limit  of flat temporal order\footnote{The MBQC-contextuality connection itself is not restricted to flat temporal order \cite{RR13}.} \cite{CohoMBQC}; (iv) computational phases of quantum matter \cite{SPT1,SPT2,SPT3,SPT4,SPT5,SPT5b,SPT6,SPT7} which relate MBQC to symmetry protected topological order.

The above facts are not isolated but form an inter-related web. For example, fact (iii) on MBQC is a consequence of fact (ii) on parity-based contextuality proofs. Furthermore, recently, a connection between the phenomenologically rather distant fields of contextual MBQCs and computational phases of matter has been established \cite{Bridge}. Cohomology emerges as a language to navigate this web of fundamental facts about quantum computation, and to explain those facts in a unified fashion.

The purpose of this paper is to extend this `web of cohomology' to a further scheme of quantum computation, namely quantum computation with magic states (QCM) \cite{BK}. In QCM,  the gate operations are reduced from a universal set to the so-called Clifford gates, which form a finite group for any number of qudits. By the Gottesman-Knill theorem, the reduced gate set alone cannot generate a quantum speedup. To retain the quantum computational power of the circuit model, the Clifford operations are supplemented by so-called `magic' states.---Which properties must those magic states have to enable universality and quantum speedup?

It has been found that discrete Wigner functions and related quasiprobability functions are indicators of quantumness for QCM \cite{NegWi}. A quantum speedup can exist only if those functions take negative values on the magic states. We show that, depending on local Hilbert space dimension and number of qudits, whether or not a Wigner function exists that can serve as an indicator of quantumness in QCM is a question of cohomology.

For odd-dimensional qudits, the phenomenology is well understood: A precondition for quantum speedup is negativity in the Wigner function \cite{Wigner1932,Woott} of the magic states \cite{NegWi,Mari}, as well as their contextuality \cite{Howard}. At a technical level, the first result rests on two facts about Gross' Wigner function \cite{GrossThesis,Gross2006}, namely (a) that it is covariant under all Clifford transformations, and (b) that its positivity is preserved under all Pauli measurements. The second result rests on the first, and further the fact that (c) non-contextual value assignments for Pauli observables exist in odd dimension.  Our discussion will reveal that the facts (a), (b), (c) are cohomological. 

Our greater phenomenological interest is with the trickier case of {\em{even}} local dimension (qubits etc.); see  \cite{Cormick,PashayanBartlett2015,KociaLove2017,ReWi,QuWi17,QuWi19,Zhu,RoM,KirbyLove, Schmid, LambdaHeim, ZurelRaussendorf2020}. Do the above results on Wigner function negativity and state-dependent contextuality cary over?---Mermin's square prevents the latter, leaving the former for discussion.  Indeed, it turns out that negativity as precondition for speedup can be reproduced in even dimension {\em{if}} one admits more general quasi-probability functions; specifically quasiprobability functions that do not stem from an operator basis and are not even unique \cite{RoM,QuWi19}. If one is not prepared to consider such generalizations, one is confronted with no-go theorems. For example, it is known that Wigner functions satisfying the assumption of diagram preservation cannot represent the stabilizer sub-theory of quantum mechanics positively \cite{Schmid}. Furthermore, for qubits, it has been shown that Clifford-covariant Wigner functions from operator bases do not exist \cite{Zhu}.\smallskip

In this paper, we extend the known no-go results for Wigner functions derived from operator bases to all even dimensions, and also to the question of positivity preservation under Pauli measurement. The latter is central for QCM; see Section~\ref{vari}. Our technical contributions are two-fold: First, we formalize the obstructions to the existence of Wigner functions with the above ``nice'' properties (a) and (b). We demonstrate that these obstructions are cohomological in nature. Second, we apply these general results to the case of even dimension.  We show that, in all even dimensions, Wigner functions constructed from operator bases cannot be Clifford covariant and cannot represent Pauli measurement positively.  Our  main results are stated as Theorems~\ref{C1b-cov} -- \ref{PosReven}.\smallskip

\noindent
\textbf{Outline.} In Section~\ref{BG} we provide the necessary background on the Pauli and Clifford groups, QCM, Wigner functions, cohomology, and contextuality. In Section~\ref{Wifu} we define the Wigner functions of present interest. Section~\ref{CliCov} is on the possibility of Clifford covariance of Wigner functions constructed from operator bases. Theorem~\ref{C1b-cov} identifies a necessary and sufficient cohomological condition for the existence of Clifford-covariant Wigner functions, and Theorem~\ref{WiCovEven} applies this condition to the case of even dimension. Section~\ref{PRPM} discusses positive representation of Pauli measurement by Wigner functions that are constructed from operator bases and fulfill the Stratonovich-Weyl criteria~\cite{Stratonovich1956,BrifMann1998}. Theorem~\ref{M_PosRep} provides a necessary and sufficient cohomological criterion for the existence of Wigner functions that represent Pauli measurement positively, and Theorem~\ref{PosReven} applies this criterion to even dimension. Section~\ref{Disc} is the discussion. 

\section{Background}\label{BG}

In this section we review the necessary background material, namely the Pauli and the Clifford group, quantum computation with magic states (QCM), Wigner functions and the Stratonovich-Weyl criteria, and cohomology. QCM is the computational model of interest for this paper, Clifford gates and Pauli measurements are the operational primitives thereof, and suitably defined Wigner functions serve as indicators or quantumness. Finally, cohomological properties determine whether the `suitable' Wigner functions {\em{can}} be defined.

We also review two types of proofs for the contextuality of quantum mechanics, namely parity-based and symmetry-based contextuality proofs. While they are not necessary to understand our main results, Theorems~\ref{C1b-cov}--\ref{PosReven}; they are based on the same cohomological structures, and thus connect the present discussion to the broader picture.

\subsection{The Pauli group and the Clifford group}

Pauli observables and Clifford unitaries are of central importance for this paper. Here we provide the definitions for reference.

\subsubsection{The Pauli group}\label{Paul}

\paragraph{Definition.} Let $d\geq 2$ be a natural number. The $1$-qudit Pauli group is defined using the usual shift $X$ and the phase $Z$ operators acting on $\CC^d$:
\begin{equation}\label{eq:1quditPauli}
X|k\rangle = |k+1\rangle, \;\;\;\; Z|k\rangle = \omega^k |k\rangle  
\end{equation}
where $k\in \ZZ_d$ and $\omega = e^{2\pi i/d}$.
Tensor products of these operators  are used to construct the $n$-qudit Pauli group. oe is a distinction between the odd- and the even-dimensional cases: Let $\mu = \omega$ and $Z_\mu=\ZZ_d$ ($\mu=\sqrt{\omega}$ and $Z_\mu=\ZZ_{2d}$) if $d$ is odd (even). Pauli operators are defined by
\begin{equation}\label{eq:Paulis}
T_a =  \mu^{\gamma(a)}Z(a_z)X(a_x)
\end{equation}
where  $a=(a_Z,a_X)\in\ZZ_d^n\times \ZZ_d^n=:E$, $Z(a_Z):=\bigotimes_{k=1}^nZ^{a_Z[k]}$, $X(a_X):=\bigotimes_{k=1}^nX^{a_X[k]}$, and $\gamma:E\to Z_\mu$ is a function  chosen such that all operators $T_a$ satisfy $(T_a)^d=I$. 
For even $d$, the requirement that
$(T_a)^d=I$ 
restricts $\gamma$ to $\gamma(a) \mod 2 \equiv(a_z)^Ta_x\mod2$.

The $n$-qudit Pauli group $P_n$ is generated by the operators $T_a$ where $a\in E$, yielding $P_n=\{\mu^\lambda T_a:\lambda\in Z_\mu, a\in E\}$.  The commutation relation among Pauli operators can be expressed in terms of a symplectic form, $T_aT_b = \omega^{[a,b]} T_bT_a$, with
\begin{equation}\label{eq:commutation}
[a,b] := (a_Z)^T b_X - (a_X)^Tb_Z \mod d.
\end{equation}
Thus, $T_a$ and $T_b$ commute if and only if $[a,b]=0$. 
Of interest to us is the multiplication table of Pauli observables, especially among commuting ones,
\begin{equation}\label{3T}
T_aT_b = \omega^{\beta(a,b)} T_{a+b},\;\; [a,b]=0.
\end{equation}
This relation defines the function $\beta$ with values in $\mathbb{Z}_d$. 

\paragraph{Structural properties.}  We have the following structural results about the multiplication table of Eq.~(\ref{3T}).

\begin{Obs}[\cite{GrossThesis}]\label{BetaOdd}
If the dimension $d$ is odd, then for any number $n$ of qudits the phases $\gamma(a)$ in Eq.~(\ref{eq:Paulis}) can be chosen such that $\beta\equiv 0$.
\end{Obs}
The proof of Observation~\ref{BetaOdd} is constructive; choose
\begin{equation} \label{PrefGamma}
\gamma(a) = -2^{-1}(a_Z)^T a_X.
\end{equation}
Among the phenomenological implications of Observation~\ref{BetaOdd} are (i) the fact that there is no parity-based contextuality proof on Pauli observables \cite{Entropy,QuWi17} (i.e., no counterpart to Mermin's square and star) when $d$ is odd; and (ii) the multi-qudit Wigner function defined in \cite{GrossThesis} is positivity-preserving under all Pauli measurements. Both properties are important for identifying Wigner function negativity and state-dependent contextuality as preconditions for quantum speedup in QCM, see \cite{NegWi},\cite{Howard}.\smallskip

The even-dimensional counterpart of Observation~\ref{BetaOdd} will be discussed in Section~\ref{sec:PauliPosEvenOdd}. For the special case of $d=2$ the following is known.
\begin{Obs}[\cite{Merm}]\label{Beta2}
If $d=2$ then for all $n\geq 2$ and any choice of the function $\gamma$ in Eq.~(\ref{eq:Paulis}) it holds that $\beta \not\equiv 0$.
\end{Obs}
The Proof of Observation~\ref{Beta2} is provided by Mermin's square \cite{Merm}.

\subsubsection{Clifford group}  

The $n$-qudit Clifford group $\text{Cl}_n$ is the normalizer of the Pauli group $P_n$ in the unitary group $U(2^n)$, with the phases modded out,  
\begin{equation}\label{CliffDef}
\Cl_n = N(P_n)/U(1).
\end{equation}
Of central interest to us is how the Clifford group acts on Pauli observables by conjugation,
\begin{equation}\label{eq:symmetry}
	g(T_a) :=gT_ag^\dagger= \omega^{\tilde\Phi_g(a)} T_{S_ga},\;\; \forall g\in \Cl_n,\;\forall a \in E,
\end{equation}
where $S_g$ is a symplectic transformation acting on $E$, i.e. an element of $\Sp(E)$ (also denoted by $\Sp_{2n}(\ZZ_d)$). Here we are using  two observations: (1) $S_g$ is a group homomorphism $E\to E$ since $g$ respects products of Pauli operators, i.e. $g(T_a T_b)=g(T_a)g(T_b)$, and (2) $S_g$ is symplectic  since $g$ respects commutators in the sense that $[g(T_a), g(T_b)] = g([T_a,T_b])$.
The Pauli group with the phases modded out, ${\cal{P}}_n:=P_n/(U(1)\cap P_n) \cong\mathbb{Z}_d^n\times \mathbb{Z}_d^n$, is a normal subgroup of $\Cl_n$, and it  holds that $\Sp(E)\cong \Cl_n/{\cal{P}}_n$.  

The phase function $\tilde{\Phi}$ plays an important role in the subsequent discussion.  $U(1)$-phases in $N(P_n)$ do not affect $\tilde{\Phi}$ in Eq.~(\ref{eq:symmetry}), which is why we  may mod out those phases to begin with, cf. the definition of $\Cl_n$ in Eq.~(\ref{CliffDef}). 

\subsection{Quantum computation with magic states}

Quantum computation with magic states (QCM) \cite{BK} is a scheme for universal quantum computation. It is closely related to the circuit model, but there is an important difference: the set of realizable quantum gates is restricted to the Clifford gates, hence not universal. Quantum computational universality is restored by the inclusion of so-called magic states.

\subsubsection{Operations in QCM}

There are two types of operations in QCM, the ``free'' operations and the resources. The free operations are (i) preparation of all stabilizer states, (ii) all Clifford unitaries, and (iii) measurement of all Pauli observables.

The resource are arbitrarily many copies of the state
\begin{equation}\label{MagStat}
|{\cal{T}}\rangle = \frac{|0\rangle + e^{i\pi/4}|1\rangle}{\sqrt{2}}.
\end{equation}
The state $|{\cal{T}}\rangle$ is called a ``magic state'', because of its capability to restore universality, given Clifford gates and Pauli measurements.

The distinction between free operations and resources in QCM is motivated by the Gottesman-Knill theorem. Namely, the free operations alone are not universal for quantum computation, and, in fact, can be efficiently classically simulated~\cite{GomaSM,AaronsonGottesman2004}. 

\subsubsection{Computational universality}

For the case of qubits, $d=2$, it is well known \cite{YaoSM} that the gates $\{\text{CNOT}_{ij}, H_i, {\cal{T}}_i;\; 1\leq i,j \leq n, i\neq j\}$ form a universal set, i.e., enable universal quantum computation. Therein, the only non-Clifford element is
${\cal{T}}_i = \exp \left(-i\frac{\pi}{8} Z_i\right)$.
This gate can be simulated by the use of a single magic state $|{\cal{T}}\rangle$ in a circuit of Clifford gates and Pauli measurements (circuit reproduced from Fig. 10.25 of \cite{NieChaSM}),
$$
\parbox{5.1cm}{\includegraphics[width=5cm]{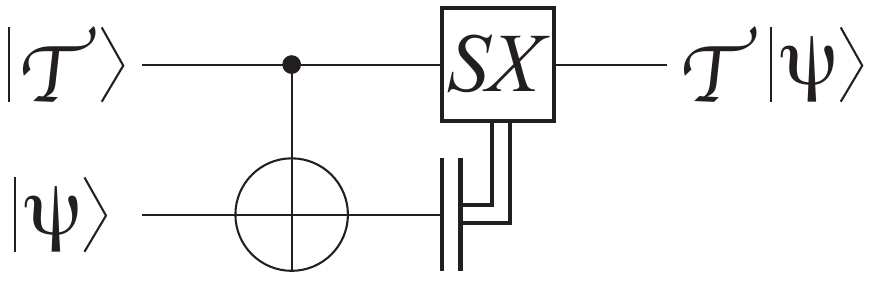}}.
$$ 
The lower qubit is measured in the $Z$-basis, and $S$ is the Clifford gate $S_i=\exp \left(-i\frac{\pi}{4} Z_i\right)$.
Thus, the magic states Eq.~(\ref{MagStat}) boost the free operations to quantum computational universality.

\subsubsection{A variant of QCM}\label{vari}

We observe that in QCM, the Clifford unitaries can be eliminated without loss of computational power~\cite{ReWi,BravyiSmolin2016}. Given the magic states, the computational power rests with the Pauli measurements.

This can be seen as follows. Consider the most general QCM, consisting of a sequence of Clifford gates interspersed with Pauli measurements, both potentially conditional on the outcomes of prior Pauli measurements. Now, starting with the last and ending with the first, the Clifford unitaries may be propagated forward in time, past the last Pauli measurement. Since the computation ends with the last measurement (all measurement outcomes have been gathered), after propagation the unitaries may be dropped without loss.

The only effect of the Clifford unitaries is that, in propagation, they change the measured observables by conjugation. But, by the very definition of the Clifford group, Pauli measurements remain Pauli measurements under conjugation by Clifford unitaries. Thus, for every QCM circuit consisting of Clifford unitaries and Pauli measurements, there is a computationally equivalent circuit that consists of Pauli measurements only.\medskip

This observation impacts the interpretation of the results of this paper. Theorems~\ref{C1b-cov} and \ref{WiCovEven} below deal with the question of when Clifford-covariant Wigner functions exist, and Theorems~\ref{M_PosRep} and \ref{PosReven} with the question of when Wigner functions exist that represent Pauli measurements positively. With the observation just made, the latter two are more important for QCM. However, we still address Clifford covariance, as it has traditionally been invoked in the discussion of QCM \cite{NegWi}, and as it is of general interest. 

\subsection{Wigner functions}\label{Wigner}

The Wigner function~\cite{Wigner1932} forms the basis of an alternative formulation of quantum mechanics. In this formalism, a linear operator $Y$ is represented by a Wigner function $W_Y$ over the position-momentum phase space. The Wigner function of a state is a quasiprobability distribution, which behaves much like a probability distribution over phase space, the basis of classical statistical mechanics. The essential difference is that the Wigner function can take negative values. This property allows the Wigner function to represent quantum mechanics, and as a result, negativity in the Wigner functions of states has been proposed as a measure that distinguishes classically behaving subsystems of quantum mechanics from those which are genuinely quantum~\cite{Hudson,KenfackZyckowski2004}.

Many other quasiprobability representations of quantum mechanics have also been defined. They are related through the Stratonovich-Weyl (SW) correspondence---a set of criteria that reasonable quasiprobability representations over generalized phase spaces should satisfy~\cite{Stratonovich1956} (also see~\cite{BrifMann1998}). Functions in the SW class have the form $W_{Y}:\mathcal{V}\rightarrow\mathbb{C}$, where $W_{Y}$ represents the linear operator $Y$ over phase space $\mathcal{V}$, and $W_\rho$ of a state $\rho$ is a quasiprobability distribution. In general, observables are represented by a dual object, the effect function $\Theta$, which maps linear operators $Y$ onto $\Theta_Y: {\cal{V}} \longrightarrow \mathbb{C}$.\footnote{The original Wigner function is, up to normalization, self dual.}

The SW criteria are as follows:
\begin{enumerate}
	\item[\hypertarget{sw0}{(sw0)}] (Linearity): the map $Y\rightarrowtail W_Y$ is one-to-one and linear,
	\item[\hypertarget{sw1}{(sw1)}] (Reality): $$W_{Y^\dagger}(x)=\left(W_Y(x)\right)^*\quad\forall x\in \mathcal{V},$$		
	\item[\hypertarget{sw2}{(sw2)}] (Standardization): $$\int_{\mathcal{V}}d\mu(x)W_Y(x)=\Tr(Y),$$
	\item[\hypertarget{sw3}{(sw3)}] (Covariance): $$W_{g\cdot Y}\left(g\cdot x\right)=W_Y\left(x\right)\quad\forall x\in\mathcal{V}\;\forall g\in G$$ where $G$ is a symmetry group of the phase space.
	\item[\hypertarget{sw4}{(sw4)}] (Traciality): $$\int_{\mathcal{V}}d\mu(x)W_{Y_1}(x)\Theta_{Y_2}(x)=\Tr(Y_1Y_2).$$
\end{enumerate}
The purpose of the traciality condition \hyperlink{sw4}{(sw4)} is to represent the Born rule in a phase-space formulation of quantum mechanics.

The interpretation of negativity in quasiprobability functions as an indicator of genuine quantumness, with many applications in quantum information processing, has also been proposed for discrete Wigner functions---quasiprobability functions used for describing finite-dimensional quantum mechanics.  For a system of $n$ qudits, each with local Hilbert space dimension $d$, the discrete Wigner function is usually defined over the finite phase space $\mathbb{F}_d^{2n}$~\cite{Woott} or $\mathbb{Z}_d^{2n}$~\cite{GrossThesis,Gross2006}, where $\mathbb{F}_d$ is the field with $d$ elements and $\mathbb{Z}_d$ is the integers mod $d$.  When $d=p$ is prime, $\mathbb{F}_d\cong\mathbb{Z}_d$, and so the two choices for the phase space are equivalent.  The scope of the former choice is limited to the case where the local Hilbert space dimension $d=p^N$ is a power of a prime since a finite field with $d$ elements exists if and only if $d$ is a power of a prime.  In this case, there is a map $\iota:\mathbb{F}_{p^N}^2\rightarrow\mathbb{F}_p^{2N}$ which preserves the structure of the phase space, and so Wigner functions defined over $\mathbb{F}_{p^N}^{2n}$ coincide with Wigner functions defined over $\mathbb{Z}_p^{2nN}$ up to relabeling of phase space points~\cite{Gross2006}.  Therefore, choosing the phase space defined over $\mathbb{F}_{p^N}$ is equivalent to representing each $d=p^N$-dimensional qudit as a system of $N$ independent $p$-dimensional qudits with the overall phase space $\mathbb{Z}_p^{2nN}$.  Many other phase spaces for finite-dimensional systems have also been proposed, see \cite{Ferrie2011} for a review.  For the purposes of the present paper we are interested in Wigner functions defined over the phase space $V:=\mathbb{Z}_d^{2n}$.

One particularly useful example of a discrete Wigner function that satisfies the SW criteria is Gross' Wigner function for systems of odd-dimensional qudits~\cite{GrossThesis}.  This is a Wigner function defined over the phase space $V=\mathbb{Z}_d^{2n}$ where $d$ is odd.  To start we choose the phase convention of the Pauli operators in Eq.~(\ref{eq:Paulis}) as in Eq.~(\ref{PrefGamma}). Then points in phase space are associated with phase space point operators defined as
\begin{equation}\label{eq:Gross}
	A_u=\frac{1}{d^n}\sum\limits_{v\in E}\omega^{-[u,v]}T_v^\dagger,\quad\forall u\in V.
\end{equation}
These operators form an orthonormal basis for the space of Hermitian operators on $d^n$-dimensional Hilbert space.  Therefore, for any density matrix $\rho$ representing a quantum state, there is a decomposition in phase point operators of the form
\begin{equation}
	\rho=\sum\limits_{u\in V}W_\rho(u)A_u.
\end{equation}
The coefficients in this expansion define the Wigner function $W_\rho:V\rightarrow\mathbb{R}$ of the state $\rho$.  Equivalently, by orthogonality, the Wigner function can be defined as
\begin{equation*}
	W_\rho(u)=\frac{1}{d^n}\Tr(A_u\rho).
\end{equation*}

This Wigner function has several properties which make it useful for describing QCM.  First, it is covariant with respect to all Clifford-group operations. That is, for any linear operator $Y$ and for all $g\in \text{Cl}_n$ it holds that
\begin{equation}\label{WigCovIntro}
W_{g(Y)}(S_g v + a_g) = W_Y(v).
\end{equation}
Therein, $S_g$ is the symplectic matrix introduced in Eq.~\eqref{eq:symmetry} and $a_g$ is a translation vector, both dependent on $g$. This follows from the fact that phase point operators map to phase point operators under conjugation by Clifford-group elements:
\begin{equation*}
	g(A_u)=A_{S_gu+a_g}\quad\forall u\in V\;\forall g\in\Cl_n.
\end{equation*}
The Wigner function also satisfies the condition of positivity preservation of the Wigner function under Pauli measurements.  That is, if a state $\rho$ has a non-negative Wigner function and a Pauli measurement is performed, the resulting postmeasurement state also has a non-negative Wigner function.  This follows from the fact that under Pauli measurements phase point operators map to probabilistic combinations of phase point operators.

In fact, the update of the phase point operators under Clifford unitaries and Pauli measurements can be computed efficiently classically.  This leads to an efficient classical simulation algorithm for QCM based on sampling from the Wigner probability distribution of the input state that applies whenever the Wigner function of the input state is non-negative.  This can be formalized in the following theorem.
\begin{Theorem}[\cite{NegWi}, adapted]\label{Start}
	Consider QCM on $n$ qudits of odd prime dimension $d$. If the initial magic state $\rho = \rho(1)_1\otimes \rho(2)_2 \otimes ..\otimes \rho(n)_n$ satisfies $W_\rho \geq 0$, then any quantum computation in the magic state model starting with $\rho$ can be efficiently classically simulated.
\end{Theorem}
We have rephrased this theorem here to make the statement self-contained. For the original formulation, see Theorem~1 in \cite{NegWi}.  This result can be extended to apply to QCM on qudits of any odd dimension~\cite{MichaelThesis}.  When the discrete Wigner function of the input state takes negative values classical simulation is still possible~\cite{PashayanBartlett2015}, but it is inefficient in general.

Many no-go results have provided obstructions to a similar result for systems of even-dimensional qudits.  For example, it has been shown that no Wigner function in the SW class for systems of multiple qubits can be Clifford covariant~\cite{Zhu}.  Quasiprobability representations for systems of even-dimensional qudits have been defined for the purpose of describing QCM, e.g.~\cite{RoM,QuWi19,KociaLove2017,KociaLove2018,ZurelRaussendorf2020}, but these generally require relaxing some of the constraints provided by the SW criteria.  In this paper we explore the underlying reasons for the difference between Wigner functions in the SW class of even- and odd-dimensional qudits in terms of Clifford covariance and positivity preservation under Pauli measurements.

\subsection{Cohomology}\label{CCC}

The purpose of the section is to explain that the function $\beta$ defined in Eq.~(\ref{3T}) and the phase function $\tilde{\Phi}$ defined in Eq.~(\ref{eq:symmetry}) are cohomological objects. As it will turn out, $\beta$ governs the existence of Wigner functions that represent Pauli measurement positively, and $\tilde{\Phi}$ governs the existence of Wigner functions that are Clifford covariant. Thus, Clifford covariance of Wigner functions and positive representation of Pauli measurement by Wigner functions are cohomological properties of the Clifford and Pauli groups. Also see Ref.~\cite{Coho} for a more detailed exposition.

\subsubsection{Motivation} 

Here we lay out a short path to recognizing $\beta$ and $\tilde{\Phi}$ as cohomological objects. A problem gets us started. The Pauli operators contain in their definition an arbitrary phase $\gamma$, cf. Eq.~(\ref{eq:Paulis}), with no physical significance.  Respecting the constraint on $\gamma$ imposed by the condition $(T_a)^d = I$, $\forall a\in E$, the phases $\gamma$ can be changed by any $\nu:E\to \mathbb{Z}_d$ as
\begin{equation}\label{trans}
\begin{array}{rcll}
\gamma &\longmapsto & \gamma +  \nu,&\text{if}\; d\; \text{is odd}, \\
\gamma &\longmapsto &\gamma + 2 \nu,& \text{if}\; d\; \text{is even}. 
\end{array}
\end{equation}
The corresponding effect on $\beta$ is, irrespective of whether $d$ is even or odd,
\begin{equation}\label{betaEquiv}
\beta(a,b) \longrightarrow \beta(a,b) + \nu(a) +\nu(b) -\nu(a+b) \mod d.
\end{equation}
An object can be of physical significance only if it is invariant under the gauge transformations of Eq.~(\ref{trans}). Can we construct such objects out of the function $\beta$?

The prepared eye recognizes Eq.~(\ref{betaEquiv}) as a cohomological equivalence transformation $\beta\longrightarrow \beta +d\nu$, where $d\nu: (a,b)\mapsto\nu(a)+\nu(b)-\nu(a+b)$ is the coboundary of $\nu$. This gives a clue: the cohomology classes $[\beta]$ are invariant objects! 

Indeed, as we demonstrate in Section~\ref{PRPM}, those cohomology classes determine the existence of Wigner functions that represent  Pauli measurement positively. To provide the foundation for the cohomological formulation, in Section~\ref{betacoc} we describe the chain complex where $\beta$ lives. \smallskip

The motivation for the cohomological description of $\tilde{\Phi}$ is analogous. With Eq.~(\ref{eq:symmetry}), the effect of the equivalence transformation Eq.~(\ref{trans})  on $\tilde{\Phi}$ is
\begin{equation}\label{PhiTrans}
\tilde{\Phi}_g(a) \longrightarrow \tilde{\Phi}_g(a) +\nu(a) - \nu(S_ga) \mod d.
\end{equation} 
Again, this looks like an equivalence transformation in group cohomology, $\tilde{\Phi} \longrightarrow \tilde{\Phi} -d^h\nu$. Therein, $d^h$ is the coboundary operator in group cohomology. The bi-complex where $\tilde{\Phi}$ lives is introduced in Section~\ref{Phicoc}. As we demonstrate in Section~\ref{CliCov}, the existence of Clifford-covariant Wigner functions hinges on a cohomological invariant extracted from $\tilde{\Phi}$.

\subsubsection{\texorpdfstring{$\beta$}{β} is a cocycle}\label{betacoc}

Let $\cC_*=(C_0,C_1,C_2,C_3)$ denote the chain complex for which $C_{k}$ is defined to be the free $\ZZ_d$-module with basis $[v_1|v_2|\cdots|v_k]$, where $v_i\in E$ and 
\begin{equation}\label{ComCon}
[v_i,v_j]=0,\; \forall i,j=1,\cdots,k.
\end{equation}
Note that $C_0=\ZZ_d$. The boundary map $\partial$ is given by the formula
\begin{equation}\label{eq:boundary}
\partial[v_1|v_2|\cdots|v_k] = [v_2|\cdots|v_k] +  \left( \sum_{i=1}^{k-1} (-1)^i[v_1|\cdots|v_i+v_{i+1}|\cdots| v_k] \right)+ (-1)^k [v_1|\cdots|v_{k-1}].
\end{equation}  
There is also an associated cochain complex $\cC^*$ whose $k$-cochains $C^k$ are given by $\ZZ_d$-module maps (i.e. $\ZZ_d$-linear  functions) $f:C_k\to \ZZ_d$. The coboundary map $\delta$ is defined by the formula $\delta f(-)=f(\partial -)$. \medskip 

With the identification $\beta([a|b]):=\beta(a,b)$, the above definitions make $\beta$ a 2-cochain in $\cC^*$, i.e. $\beta \in C^2$. But we can say more; $\beta$ is in fact a cocycle, $d\beta =0$. Namely, associativity of operator multiplication, $T_a(T_bT_c) = (T_aT_b)T_c$, implies, for all $a,b,c \in E$ such that $[a,b]=[a,c]=[b,c]=0$, 
$$
\beta(b,c) -\beta(a+b,c)+\beta(a,b+c)-\beta(a,b)=0.
$$
The surface 
$F:=[b|c]-[a+b|c]+[a|b+c]-[a|b]$
 is the boundary of the volume 
 $V=[a|b|c]$, $F=\partial V$. Thus, $0=\beta(\partial V)= d\beta (V)$, for all volumes $V$. Hence, 
\begin{equation}
d\beta =0,
\end{equation}
as claimed. The equivalence class of $\beta$ is now defined by 
\begin{equation}\label{betaClass}
[\beta]:= \{\beta + d\nu, \nu \in C^1\}. 
\end{equation}
The equivalence classes of $2$-cocycles form the second cohomology group $H^2(\cC,\mathbb{Z}_d)$. They are independent of the choice of phase $\gamma$ in Eq.~(\ref{eq:Paulis}).\smallskip

To provide a first application of the cohomological formulation, recall that in Section~\ref{Paul} we discussed whether the phase factors in the Pauli operator multiplication table of Eq.~(\ref{3T}) can be eliminated by a clever choice of the phase convention $\gamma$. We now find that this is a topological question. Namely, with Eqs.~(\ref{trans}) and (\ref{betaClass}), the phase factor $\omega^\beta$ can be removed if and only if $[\beta]=0$. Observations~\ref{BetaOdd} and \ref{Beta2} may thus be reformulated in a cohomological fashion as (i) If $d$ is odd then for any $n$ it holds that $[\beta]=0$, and (ii) If $d=2$ then for any $n\geq 2$ it holds that $[\beta]\neq 0$.

\subsubsection{The group cocycles \texorpdfstring{$\tilde{\Phi}$}{Φ-tilde} and \texorpdfstring{$\Phi_\cov$}{Φ-cov}}\label{Phicoc}

 In this section we will regard $\tilde \Phi$, introduced in Eq.~(\ref{eq:symmetry}), as a cocycle and relate its cohomology class to certain properties of Wigner functions. For this we need to introduce a chain complex which is slightly different than the one used above. The main difference is that we remove the commutativity constraint imposed on tuples constituting the basis of the chain complexes. 
We define a chain complex $\tilde \cC_*=(\tilde C_0,\tilde C_1,\tilde C_2,\tilde C_3)$. Here $\tilde C_k$ is the free $\ZZ_d$-module with basis consisting of the tuples $[v_1|v_2|\cdots|v_k]$ where $v_i\in E$. 
The boundary map is the same as in Eq.~(\ref{eq:boundary}). The associated cochain complex is denoted by $\tilde \cC^*=(\tilde C^0,\tilde C^1,\tilde C^2,\tilde C^3)$ and the coboundary map $\delta:\tilde C^k \to \tilde C^{k+1}$ is induced by the boundary map as before. Note that by definition $\tilde C_1=C_1$ and $\tilde C^1=C^1$, thus in this case we remove the extra decoration for simplicity of notation. We also note that  the (co)chain complex defined here is the standard complex (which is called the bar construction) that computes the group (co)homology of the abelian group $E$. 

For a symmetry group specified by a subgroup $G\subseteq \Cl_n$, we consider the bicomplex $C^p(G,\tilde C^q)$; see also \cite[Section 5.2]{Coho}.
The bicomplex $C^p(G,\tilde C^q)$ 
comes with two types of coboundaries: group cohomological  $d^h:C^p(G,\tilde C^q)\to C^{p+1}(G,\tilde C^q)$, and $d^v:C^p(G,\tilde C^q)\to C^{p}(G,\tilde C^{q+1})$ induced by $\delta$.  

The phase function 
$\tilde\Phi:G \to C^1$  is per definition a $1$-cochain in group cohomology. Its coboundary, a $2$-cochain, is
\begin{equation}\label{dhDef1}
(d^h\tilde{\Phi})_{g,h}(a)  :=  \tilde{\Phi}_h(a) -  \tilde{\Phi}_{gh}(a) + \tilde{\Phi}_g(S_ha).
\end{equation}
In fact, $\tilde{\Phi}$ is not only a $1$-cochain but a $1$-cocycle, i.e., $d^h\tilde{\Phi} =0$. Namely, with the associativity of matrix multiplication, it holds that $(gh)(T_a)=g(h(T_a))$, $\forall g,h \in G$, $\forall a\in E$. Evaluating both sides using Eq.~(\ref{eq:symmetry}) yields
$\tilde{\Phi}_{gh}(a) = \tilde{\Phi}_h(a) + \tilde{\Phi}_g(S_ha),\; \forall g,h\in G$ and $\forall a\in E$.
Thus, with Eq.~(\ref{dhDef1}), $(d^h\tilde{\Phi})_{g,h}(a) =0$, for all $g,h\in G$ and all $a\in E$. $\tilde{\Phi}$ is indeed a group cocycle.\smallskip

This group cocycle may be trivial or nontrivial. Consider a $0$-cochain $\nu$ in group cohomology. Its coboundary, a $1$-cocycle, is
\begin{equation}\label{dhDef0}
(d^h\nu)_g(a) := \nu(S_ga) - \nu(a).
\end{equation}
We define the group cohomology classes
$$
[\tilde{\Phi}]=\{\tilde{\Phi} + d^h\nu,\;\forall \nu \in C^1\}.
$$
A cocycle $\tilde{\Phi}$ is trivial iff it can be written in the form $\tilde{\Phi} = d^h\nu$, for some $\nu\in C^1$. A class $[\tilde{\Phi}]$ is trivial, $[\tilde{\Phi}]=0$, if and only if it contains $\tilde{\Phi}\equiv 0$.\medskip

The cocycle $\tilde{\Phi}$ may not only be evaluated on edges $a\in E$ but, by linear extension, on all $1$-chains: A $1$-chain is a $\ZZ_d$-linear sum of elements in $E$ and we define 
$$
\tilde \Phi_g(\sum_j \alpha_j [a_j]) = \sum_j \alpha_j \tilde \Phi_g( a_j), \;\; \alpha_j\in \ZZ_d,a_j\in E,
$$
where each $\tilde \Phi_g(a_j)$ is determined by Eq.~(\ref{eq:symmetry}). Subsequently, we will have occasion to evaluate $\tilde{\Phi}$ on boundaries $\partial f$, for $f \in C_2$. When $G=\Cl_n$ it is easily verified that, for boundaries $\partial f$,
$$
\tilde{\Phi}_g(\partial f) = \tilde{\Phi}_{[g]}(\partial f),\;\forall g\in \Cl_n,
$$
where $[g] \in \Cl_n/{\cal{P}}_n$. That is, when $\tilde{\Phi}$ is evaluated on a boundary, it depends on its first argument $g$ only through the equivalence class $[g]$.

To formalize this property in general, let $N\subset G$ denote the subgroup of symmetries $g$ such that $S_g$ is the identity transformation. The quotient group $Q=G/N$ is the essential part of the symmetries acting on the complex. Let $\tilde B_1$ denote the image of the boundary map $\partial : \tilde C_2 \to C_1$. We write $U_\cov$ for the set of $\ZZ_d$-module maps $\tilde B_1\to \ZZ_d$. We choose a set-theoretic section $\theta:Q\to G$ of the quotient map $\pi:G\to Q$; i.\,e., $\theta(\pi(g))\in\pi(g)$. Then $\Phi_\cov\in C^1(Q,U_\cov)$ is defined to be the composite
\begin{equation}\label{eq:Phi}
\Phi_\cov: Q \xrightarrow{\theta} G \xrightarrow{\tilde\Phi} C^1 \xrightarrow{d^v}  U_\cov
\end{equation}
where the last map can also be thought of as the restriction of a $\ZZ_d$-module map $C_1\to \ZZ_d$ to the boundaries $\tilde{B}_1$.  
More explicitly we have
$$
\Phi_\cov(q,\partial f) = d^v\tilde\Phi_{\theta(q)}(f) = \tilde\Phi_{\theta(q)}(\partial f)
$$
for any $q\in Q$ and $f\in \tilde C_2$. Although $\tilde C^1=C^1$, the object $U_\cov$ is different from its counterpart $U$ introduced in \cite[Eq.~(39)]{Coho} as part of the symmetry discussion in  contextuality. See also Theorem \ref{C1b} below.

Like $\tilde{\Phi}$, $\Phi_\cov$   is also a group cocycle, $d^h\Phi_\cov =0$. The cocycle class of $\Phi_\cov$ is given by
$$
[\Phi_\cov] = \{\Phi_\cov + d^h\nu,\; \nu \in U_\cov\}.
$$
$[\Phi_\cov]$ is the object of interest for Wigner function covariance.\smallskip

As a first application of the group cohomological formalism, recall Theorem~37 from \cite{GrossThesis}, on the structure of the Clifford group in odd dimension. Items (i) and (iii) of that theorem read: (i) For any symplectic $S$, there is a unitary operator $\theta(S)$ such that $\theta(S) T_a \theta(S)^\dagger = T_{Sa}$; (iii) Up to a phase, any Clifford operation is of the form
$U = T_b\, \theta(S)$, for a suitable $b\in E$ and symplectic transformation $S$.

Comparing the relation in item (i) with the general relation Eq.~(\ref{eq:symmetry}), we find that $\tilde{\Phi}_{\theta(S)}=0$ for all symplectic $S$, w.r.t. the phase convention $\gamma$ chosen in \cite{GrossThesis}. Thus, in particular, $\tilde{\Phi}_{\theta(S)}(\partial f)=0$ for all $f\in \tilde C_2$. Since Pauli flips don't change $\tilde{\Phi}$ on boundaries, $\tilde{\Phi}_{T_b\,\theta(S)}(\partial f)=0$ for all $f\in \tilde C_2$ and all $b\in E$. But with item (iii) this covers the entire Clifford group. I.e., there is a phase convention $\gamma$ such that $\Phi_\cov \equiv 0$. The phase-convention independent version of this statement is\smallskip

\begin{Obs}\label{PhiOdd}
For the Clifford group $Cl_n$ in any odd dimension $d$ it holds that $[\Phi_\text{\upshape{cov}}]=0$.
\end{Obs}

\subsubsection{Splitting and the group cocycle \texorpdfstring{$\zeta$}{ζ}} \label{sec:splitting-group-cocycle}

For a subgroup $G\subset \Cl_n$ define $N=G\cap {\mathcal P}_n$ as before.
Since $N$ is a normal subgroup we can consider the quotient group $Q=G/N$.
A structural question about $G$ is how it is put back together from its two parts $N$ and $Q$. A particularly simple composition is the semi-direct product, $G \cong Q \ltimes N$. If it applies, then $G$ is said to `split'. But the semi-direct product is only one among a number of ways to compose $Q$ and $N$. Those ways are classified by a cohomology group \cite{adem2013cohomology}. 

The notion of splitting is of interest here because, as we shall prove in Section~\ref{Split}, it governs the existence of the faithful group action which is required in the l.h.s. of Eq.~(\ref{WigCovIntro}) and is, thus, a precondition for covariance.

Let us formulate splitting as a cohomological property. 
We identify $Q$ with the subgroup of $\Sp(E)$ to which it is isomorphic by Eq.~\eqref{eq:symmetry}. For convenience, we denote $T_a$ modulo phase by $t_a$, such that ${\cal{P}}_n=\{t_a,a\in E\} \cong E$.
Any set-theoretic section $\theta:Q\to G$ has the following properties:
\begin{subequations}
\label{MuProp}
\begin{align}
	\theta(S) t_a (\theta(S))^\dagger &= t_{Sa}, \;\forall t_a \in E,\label{MuPropA} \\
	g &= t_{\alpha_g} \theta(S_g), \; \forall g \in G,\label{MuPropB}\\
	\theta(S_1)\theta(S_2)&=t_{\zeta(S_1,S_2)}\theta(S_1S_2),\; \zeta: Q \times Q \longrightarrow E\label{MuPropC}
\end{align}
\end{subequations}
Herein $t_{\alpha_g},t_{\zeta(S_1,S_2)}\in N$. The section $\theta$ is not unique. From any given $\theta$ we may switch to new $\theta'$ via
\begin{equation}\label{muCha}
	\theta'(S) = t(S) \theta(S),\;t(S)\in N,\;\forall S\in Q.
\end{equation}
The function $\zeta$ changes under the transformation in Eq.~(\ref{muCha}), and the set $[\zeta]$ of functions $\zeta'$ that can be reached from $\zeta$ via Eq.~(\ref{muCha}) is an element of the second cohomology group $H^2(Q,E)$. The group $G$ splits if and only if $\zeta$ vanishes for a suitable choice of $\theta$. In this case we say that $[\zeta] =0\in H^2(Q,E)$.
See Eq.~(\ref{eq:zeta}) in the appendix for a particular choice of the cocycle $\zeta$. 
\smallskip

\subsection{Contextuality}\label{Context}

Unlike the previous parts of this background section, the material on contextuality discussed here is not necessary to understand the main results of this paper, Theorems~\ref{C1b-cov} -- \ref{PosReven}. However, it is helpful for connecting to the broader picture. Namely, two previously established theorems on state-independent contextuality \cite{Coho}, restated below as Theorems~\ref{ParCon} and \ref{C1b}, are structurally akin to Theorems~\ref{C1b-cov} and \ref{M_PosRep}, on the existence of Wigner functions with ``nice'' properties. They invoke the same cohomological conditions. This background portion prepares for the discussion in Section~\ref{Disc}.\medskip  

Contextuality is a foundational property that distinguishes quantum mechanics from classical physics. A priori, one may attempt to describe quantum phenomena by so-called hidden variable models (HVMs) in which all observables have predetermined outcomes that are merely revealed upon measurement. A probability distribution over such predetermined outcomes is then intended to mimic the randomness of quantum measurement. An additional constraint on HVMs is the assumption of noncontextuality: the value assigned to any given observable just depends on that particular observable, and not on any compatible observable that may be measured in conjunction. The Kochen-Specker theorem says that, in Hilbert spaces of dimension greater than two, no noncontextual hidden variable model can reproduce the predictions of quantum mechanics. Since noncontextual hidden variable models fail in this realm, quantum mechanics is said to be ``contextual''.

The original proof of the Kochen-Specker theorem is intricate. However, when sacrificing a modest amount of generality, namely the case of Hilbert space dimension 3, a very simple proof can be given---Mermin's square~\cite{Merm}.
Here we review two types of proofs of the Kochen-Specker theorem, the parity-based proofs (Mermin's square is the simplest example), and the symmetry-based proofs. \smallskip

\paragraph{Parity-based proofs.} The main examples of parity-based proofs are the well-known Mermin's square in dimension 4 and Mermin's star in dimension 8 \cite{Merm}. In those examples, the proof is based on a cleverly chosen set of Pauli observables. However, for parity proofs to work, the observables don't need to be of Pauli type; it suffices that all their eigenvalues are $k$th roots of unity for some fixed $k\in\mathbb{N}$.  

Parity proofs have a cohomological formulation. With  the cocycle $\beta$ and corresponding cohomology class $[\beta]$ in $H^2(\cC)$ defined as in Section \ref{CCC}, but this time for a set ${\cal{O}}$ of observables whose eigenvalues are all powers of $\omega$ (not necessarily, but possibly, Pauli observables), we have the following result.

\begin{Theorem}[\cite{Coho}]\label{ParCon}
For   a set of observables $\oO$ with all eigenvalues of form $e^{i2\pi\, m/k}$, for $m,k\in \mathbb{N}$ and $k$ fixed, a parity-based contextuality proof exists if and only if $[\beta]\neq 0$.
\end{Theorem}

\paragraph{Symmetry-based proofs.} Proofs of the Kochen-Specker theorem may also be based on the transformation behavior of a set ${\cal{O}}$ of observables under a symmetry group $G$ \cite{Coho}.  To be a symmetry group, $G$ (i) needs to map the set $\oO$ to itself up to phases that preserve the constraint on the eigenvalue spectrum, and (ii) needs to preserve algebraic relations among the transformed observables. Again, the symmetry-based contextuality proofs have a cohomological formulation.

\begin{Theorem}[\cite{Coho}]\label{C1b}
For a given set ${\cal{O}}$ of observables as above, and a corresponding symmetry group $G$, if $[\Phi]\neq 0 \in H^1(Q,U)$ then ${\cal{O}}$ exhibits state-independent contextuality.
\end{Theorem} 
There is a small difference between $\Phi$ in Theorem~\ref{C1b} and $\Phi_\cov$ from Section~\ref{Phicoc}. Both phase functions are defined only on boundaries $\partial f$ where $f=[a|b]$. However, for $\Phi$ all $f$ are constraint to $[a,b]=0$, whereas for $\Phi_\cov$ as defined in Section~\ref{Phicoc} this extra condition is not imposed.

\section{Wigner functions from operator bases}\label{Wifu}

In this section we define the Wigner functions we are concerned with in this paper, namely Wigner functions based on operator bases, and derive elementary properties of them. Simultaneously, we also define matching effect functions. This lays the groundwork for Sections~\ref{CliCov} and \ref{PRPM}, where we discuss Clifford covariance of Wigner functions and positive representation of Pauli measurement, respectively.

Quasiprobability functions derived from operator bases are a natural choice: the original Wigner function \cite{Wigner1932}, its finite-dimensional adaptions \cite{Woott}, and thereof in particular Gross' Wigner function in odd dimension \cite{GrossThesis,Gross2006} are all of this form\footnote{Note that in the infinite-dimensional case of the original Wigner function, the set of phase point operators is not a basis in the standard sense, i.e., it is not a Hamel basis, but it is a Hilbert basis for the ambient space that the density operators live in viewed as an inner product space (see e.g. Ref.~\cite[\S9]{Roman2008} for details on this distinction). In finite dimensions there is no such distinction.}.

We remark that, for even local dimension $d=2$, if we move beyond operator bases and admit non-unique probability functions defined on larger (generalized) phase spaces then Theorem~\ref{Start} is known to extend \cite{RoM,QuWi19}. However, moving into this territory, we also meet {\em{probability}} functions that represent universal QCM \cite{ZurelRaussendorf2020}. No negativity is needed, neither in the states nor the operations! For those scenarios, since negativity never occurs, it cannot be a precondition for speedup.

Quasiprobability functions derived from operator bases represent a calmer and more standard scenario, and therefore we settle their case here. For the remainder of this paper, we impose the condition
\begin{itemize}
\item[\hypertarget{OB}{(OB)}]{For any linear operator $Y$, a corresponding Wigner function $W_Y$ satisfies
\begin{equation}\label{OB}
	Y = \sum_{v\in V} W_Y(v) A_v, 
\end{equation}
where the operators $\{A_v,\, v\in V\}$ form an operator basis and the phase space is $V=\ZZ_d^n \times \mathbb{Z}_d^n$.}
\end{itemize}
As before in Section~\ref{Wigner}, in addition to the quasiprobability function $W$ representing quantum states, we define a dual object, the effect function $\Theta$, representing observables. For any linear operator $Y$, $\Theta_Y: V \longrightarrow \mathbb{C}$.\smallskip

{\em{$E$ vs. $V$.}} We briefly comment on the distinction between the sets $E$ of Pauli labels and the phase space $V$. Both are isomorphic to the group $\mathbb{Z}_d^n \times \mathbb{Z}_d^n$, but they represent different physical objects. $E$, as introduced right after Eq.~(\ref{eq:Paulis}), is the set of Pauli operators modulo phase. The multiplication of Pauli operators induces an addition in $E$, endowing it with a group structure. $V$, introduced right above through Eq.~(\ref{OB}), is the set of points in phase space. Once an origin is fixed, points in phase space may also be added; but in the present paper we make no use of this operation. Finally, the property \hyperlink{SW3}{(SW3)} of Pauli covariance implies an action of $E$ on $V$, 
\begin{equation}\label{action}
E\ni a: v \mapsto a(v)=(a+v) \in V,\;\; \forall v\in V.
\end{equation}
Some of our  results refer to the SW criteria, in the following formulation adapted to finite-dimensional Hilbert spaces:
\begin{itemize}
\item[\hypertarget{SW1}{(SW1)}] (Reality):$$W_{Y^\dagger}(u)=\left(W_Y(u)\right)^*\quad\forall u\in V,$$
\item[\hypertarget{SW2}{(SW2)}]{(Standardization):$$\sum\limits_{u\in V}W_Y(u)=\Tr(Y),$$}
\item[\hypertarget{SW3}{(SW3)}] (Pauli covariance): $$W_{T_a(Y)}(u+a)=W_Y\left(u\right)\quad\forall u\in V\;\forall a\in E,$$
\item[\hypertarget{SW4}{(SW4)}] (Traciality): $$\sum\limits_{u\in V}W_{Y_1}(u)\Theta_{Y_2}(u)=\Tr(Y_1Y_2).$$
\end{itemize}
Condition \hyperlink{OB}{(OB)} implies \hyperlink{sw0}{(sw0)} but is not implied by it. \hyperlink{SW2}{(SW2)} and \hyperlink{SW4}{(SW4)} are obtained from \hyperlink{sw2}{(sw2)} and \hyperlink{sw4}{(sw4)} by choosing a natural measure $\mu$. As opposed to \hyperlink{sw3}{(sw3)}, \hyperlink{SW3}{(SW3)} refers to a specific symmetry group $G$---the Pauli group---and a particular action of the symmetry group on the phase space. 

Among the Wigner functions permitted by the constraints \hyperlink{OB}{(OB)} and \hyperlink{SW1}{(SW1)}--\hyperlink{SW4}{(SW4)}, we are interested in those that transform covariantly under all Clifford gates and represent Pauli measurement positively.
Note that we do not require diagram preservation---the assumption that the representation of a composition of processes is identical to the composition of the representations of the processes~\cite{Schmid2020}.  Full diagram preservation requires this for both parallel and sequential composition of states and operations.  In particular, for parallel composition of states, diagram preservation requires that for a product state $\rho\otimes\sigma$ we have $W_{\rho\otimes\sigma}(u\oplus v)=W_\rho(u)W_\sigma(v)$.  This is a strong assumption which, though reasonable for some local realist models, is not satisfied by several models which are useful for describing quantum computation, e.g.~\cite{RoM,LillystoneEmerson2019,QuWi19,ZurelRaussendorf2020,ZurelHeimendahl2021}. In fact, this assumption is incompatible with the ability of a model to simulate contextuality as in quantum computation on multiple qubits~\cite{KaranjaiBartlett2018}. The conditions imposed on quasiprobability distributions in~\cite{Schmid2020} include diagram preservation and imply \hyperlink{OB}{(OB)}, \hyperlink{SW1}{(SW1)}, \hyperlink{SW2}{(SW2)}, and \hyperlink{SW4}{(SW4)}~\cite{Schmid,Ferrie2008}. If a quasiprobability distribution from~\cite{Schmid2020} positively represents the entire stabilizer subtheory, it additionally fulfills \hyperlink{SW3}{(SW3)}~\cite{Schmid} and, thus, all of our conditions.
\medskip

Regarding the effect function $\Theta$, we observe that all such functions admissible by \hyperlink{OB}{(OB)} and \hyperlink{SW4}{(SW4)} take a simple form, 
\begin{equation}\label{ThetaTrace}
\Theta_Y(v)=\Tr(YA_v),\;\; \forall v\in V.
\end{equation}
Namely, for $Y=A_u$ in Eq.~(\ref{OB}) the condition \hyperlink{OB}{(OB)} implies that $W_{A_u}(v) = \delta_{uv}$. Using this in the traciality condition \hyperlink{SW4}{(SW4)} for $Y_1=A_v$ implies Eq.~(\ref{ThetaTrace}). Note furthermore that Eq.~\eqref{ThetaTrace} implies \hyperlink{SW4}{(SW4)}.

We further observe that for the projectors $\Pi_{a,s}$ corresponding to the measurement of the Pauli observable $T_a$ with outcome $s$, for all $\Theta$, $W$ satisfying  \hyperlink{OB}{(OB)},  \hyperlink{SW2}{(SW2)},  \hyperlink{SW4}{(SW4)}, it holds that
\begin{equation}\label{ThetaSum}
\sum_{s} \Theta_{\Pi_{a,s}}(v) =1,\;\; \forall v\in V. 
\end{equation}
Namely, $\sum_{s} \Theta_{\Pi_{a,s}}(v) = \sum_{s} \text{Tr}(\Pi_{a,s} A_v) = \text{Tr}(A_v)$. The last expression equals $\sum_{u \in V}W_{A_v}(u) = \sum_{u \in V}\delta_{uv} =1$, with \hyperlink{SW2}{(SW2)}, yielding Eq.~(\ref{ThetaSum}).

To prepare  for the subsequent discussion, we parametrize the phase point operators. Pauli covariance \hyperlink{SW3}{(SW3)} holds if and only if
\begin{equation}\label{PhaPoDef}
A_v = \frac{1}{d^n} \sum_b \omega^{-[v,b]}c_b T_b^\dagger, \; \forall v\in V,
\end{equation}
where $c_b\in \mathbb{C}$ for all $b$, $V=\mathbb{Z}_d^n\times \mathbb{Z}_d^n$.

Additionally imposing Standardization \hyperlink{SW2}{(SW2)} and the fact that the operators $A_v$ span an operator basis \hyperlink{OB}{(OB)} is equivalent to the following two conditions on the coefficients $c_b$ in Eq~(\ref{PhaPoDef}), respectively:
\begin{subequations}
\label{StandBas}
\begin{align}
\label{Stand}
c_0 &= \mu^{\gamma(0)},\\
\label{Bas}
c_b &\neq 0,\; \forall b.
\end{align}
\end{subequations}
With these  properties established, we are now ready to address the questions of Clifford covariance and positive representation of Pauli measurement.

\section{Clifford covariance}\label{CliCov}

In this section we establish a necessary and sufficient condition for the existence of Clifford covariance of Wigner functions constructed from operator bases, for varying local Hilbert space dimension and number of particles. There are cohomological obstructions to the existence of such Wigner functions, and in even local dimension these obstructions don't vanish.

\subsection{When are Wigner functions Clifford covariant?}

We start out from the following
\begin{Def}\label{CC} For any subgroup $G\subseteq \Cl_n$, a Wigner function $W$ is called $G$-covariant if
	\begin{equation}
		\label{CovarCond}W_{g(Y)}(S_gu+a_g)=W_Y(u)\quad\forall u\in V\;\forall g\in G
	\end{equation}
	where $S_g$ is the symplectic transformation defined in Eq.~(\ref{eq:symmetry}) and $a_g \in E$. When $G=\Cl_n$, we say that $W$ is Clifford covariant.
\end{Def}
This definition, in the most general case of full Clifford covariance, is analogous to the one given in \cite{GrossThesis} for odd dimension $d$; cf. Theorem~41 therein. From the perspective of classical simulation of QCM, the usefulness of the above covariance property is that the map $g:\; u \mapsto S_gu+a_g$ can be efficiently computed, for all $g \in \text{Cl}_n$. We remark that the addition is between an element of $V$ and one of $E$, cf. Eq.~(\ref{action}). For consistency, we observe that in the limiting case of $G=P_n$, $S_g=I$ for all $g\propto T_u \in P_n$, and Eq.~(\ref{CovarCond}) reduces to the Pauli covariance condition (\hyperlink{SW3}{SW3}).\smallskip

It is useful to restate the covariance condition Eq.~(\ref{CovarCond}) in terms of phase point operators. With Eq.~(\ref{OB}) it holds that $W_{A_w}=\delta_w(v)$. Now, for any $g\in G$ we have
\begin{equation}\label{CovarC2}
\begin{array}{rcl}
g(A_w) &=& \sum_v W_{g(A_w)}(v) A_v\\
&=& \sum_v W_{g(A_w)}(S_g v + a_g) A_{S_g v + a_g}\\
&=&\sum_v W_{A_w}(v) A_{S_g v + a_g}\\
&=&\sum_v {\delta_w}(v) A_{S_g v + a_g}\\
&=& A_{S_g w + a_g}
\end{array}
\end{equation}
Therein, in the second line we have relabeled the summation index, and in the third line we have used the covariance condition Eq.~(\ref{CovarCond}). Thus, Eq.~(\ref{CovarCond}) implies Eq.~(\ref{CovarC2}).

Now we show the reverse. Assuming Eq.~(\ref{CovarC2}), the operator $g(Y)$ can be expanded in two ways
$$
g(Y) = \sum_v W_Y(v) g(A_v) =  \sum_v W_Y(v) A_{S_gv+a_g},
$$
and
$$
g(Y) = \sum_v W_{g(Y)}(v) A_v = \sum_v W_{g(Y)}(S_gv+a_g) A_{S_gv+a_g}
$$
Comparing the r.h.s.-es, and noting that by \hyperlink{OB}{(OB)} the expansions are unique, Eq.~(\ref{CovarCond}) follows. 

Both arguments combined show that a Wigner function $W$ satisfying \hyperlink{OB}{(OB)} is $G$-covariant if and only if
\begin{equation}\label{CovarC3}
	g(A_v) = A_{S_g v + a_g},\;\; \forall v\in V,\; \forall g\in G.
\end{equation}

\subsection{Existence of Clifford-covariant Wigner functions}\label{CliffEx}

Given a particular Wigner function in terms of phase point operators, Clifford covariance may be verified using the criterion Eq.~(\ref{CovarC3}). Now we are concerned with the question of whether for a system of $n$ qudits of $d$ levels a Clifford-covariant Wigner function {\em{exists}}. The result of this section is a cohomological criterion for the existence of a Clifford-covariant Wigner function.

\begin{Theorem}\label{C1b-cov}
	For any integers $n\geq 1$, $d\geq 2$, a Clifford-covariant Wigner function according to \hyperlink{OB}{(OB)} exists if and only if  $[\Phi_{\text{\upshape cov}}] =  0 \in H^1(Q,U_{\text{\upshape cov}})$.
\end{Theorem}  
{\em{Proof of Theorem~\ref{C1b-cov}.}} ``Only if'': Suppose $W$ is Clifford covariant. Then, with Eq.~(\ref{CovarC3}), $T_a(A_v)=A_{v+y(a)}$, $\forall a\in E$, where for readability we write $y(a)$ instead of $a_g$ with $g=T_a$.  First, we want to show that the function $y: E \cong \mathbb{Z}_d^{2n}\rightarrow V\cong \mathbb{Z}_d^{2n}$ is invertible. $T_aT_b(A_v)=T_{a+b}(A_v)$ implies that $y$ is linear. Therefore, $y$ is invertible if and only if its kernel is trivial, $\operatorname{Ker}(y)=\{0\}$. We expand $A_v=\sum_{b \in E} c_b(v)T_b^\dagger$. Note that, in contrast to Ansatz~\eqref{PhaPoDef}, this expansion does not restrict $A_v$. Now assume that $\operatorname{Ker}(y)\ni a \neq 0$. Then $\sum_b c_b(v)T_b^\dagger=A_v=T_a(A_v)=\sum_b c_b(v) \omega^{[b,a]}T_b^\dagger$ $\forall v$ or, equivalently, 
$$
c_b(v)=0 \;\;\forall b\in E: [b,a]\neq 0,\;\forall v\in V.
$$
For any $a\neq 0$ there is a $b \in E$ such that $[b,a]\neq 0$, and hence a $b\in E$ such that $c_b(v)=0$, $\forall v$. Since this contradicts \hyperlink{OB}{(OB)}, $y$ must be invertible.  We may thus write
\begin{equation}\label{eq:y}
T_{y^{-1}(w)}(A_0) = A_w,\; \forall w \in V.
\end{equation}
Because of Clifford covariance, for any $g\in \text{Cl}_n$ it holds that $g(A_0^\dagger) = A_{a_g}^\dagger$. Therein, we have $g( A_{0}^\dagger) = \sum_b c_b^*(0)\, g(T_b) =  \sum_b c_b^*(0) \omega^{\tilde{\Phi}_g(b)} \,T_{S_gb}$, 
and, with Eq.~(\ref{eq:y}), $A_{a_g}^\dagger = \left(T_{y^{-1}(a_g)}(A_0) \right)^\dagger  = \sum_b \omega^{[y^{-1}(a_g),b]} c_b^*(0)\, T_b$. For any $g\in \text{Cl}_n$ and any pair $b,S_gb\in E$ it follows that
\begin{equation}\label{Orbit}
c_{S_gb}(0) = c_b(0)\, \omega^{[y^{-1}(a_g),S_gb]-\tilde{\Phi}_g(b)}.
\end{equation}
Thus, for the entire Clifford orbit $\langle b\rangle$ of any $b\in E$, the expansion coefficients $c_b(0)$ have the same magnitude, and differ only by phase factors $\omega^m$, $m \in \mathbb{N}$. They can thus be written in the form
\begin{equation}\label{Orbit2}
c_b(0) = c_{\langle b \rangle} \,\omega^{\nu(b)}, \forall b\in E,
\end{equation} 
where $c_{\langle b\rangle} \in \mathbb{C}$ only depends on the Clifford orbit $\langle b\rangle$ of $b$, and $\nu: E \longrightarrow \mathbb{Z}_d$. 

Further inspecting Eq.~(\ref{Orbit}), the l.h.s. depends on $g$ only through $S_gb\in E$, whereas the r.h.s. has the a priori more general dependence through $\tilde{\Phi}_g$. Comparing Eqs.~(\ref{Orbit}) and (\ref{Orbit2}), we find that the functions $\nu$ and $\tilde{\Phi}$ are mutually constrained by the relation
$$
\nu(S_gb)-\nu(b) = [y^{-1}(a_g),S_gb] - \tilde{\Phi}_g(b),\;\; \forall b\in E,\; \forall g\in \text{Cl}_n.
$$
Now consider a face $f \in \tilde{C}_2$ with boundary $\partial f  =[a]+[b]-[a+b]$, and add up the above relations for the edges in the boundary. Because of its linearity, the commutator term then vanishes, and we obtain
$$
\nu(g\partial f) - \nu(\partial f) = - \tilde{\Phi}_g(\partial f) = -\Phi_\cov([g],\partial f),\;\forall f \in C_2({\cal{C}}'), \;\forall g\in \text{Cl}_n,
$$
where $g\partial f:=[S_ga]+[S_gb]-[S_g(a+b)]$. Thus, $\Phi_\cov = d^h (-\nu)$, i.e., $[\Phi_\cov] = 0$.\medskip

``If'': Assume that $[\Phi_\cov]=0$. Thus, there is a phase convention $\gamma$ such that $\tilde{\Phi}^{(\gamma)}|_{\tilde B_1}\equiv 0$. But this means that $\tilde{\Phi}^{(\gamma)}_g(\cdot)$ is a linear function for any $g\in \text{Cl}_n$, and we may write it as
$$
\tilde{\Phi}^{(\gamma)}_g(a) = [x_g,a],
$$
for suitable $x_g$, $g\in \text{Cl}_n$. Let $W^{(\gamma)}$ be the Wigner function defined by the phase point operators 
$
A^{(\gamma)}_v = \frac{1}{d^n} \sum_b \omega^{-[v,b]} \left(T_b^{(\gamma)}\right)^\dagger.
$
Since the $A^{(\gamma)}_v$ constitute a special case of Ansatz~\eqref{PhaPoDef} with Condition~\eqref{Bas}, we already know that they span an operator basis \hyperlink{OB}{(OB)}. Moreover, $W^{(\gamma)}$ is Clifford covariant. Namely,
$$
\begin{array}{rcl}
g\!\left(A_v^{(\gamma)}\right) &=&  \frac{1}{d^n} \sum_b \omega^{-[v,b]} \, g\!\left[\left(T_b^{(\gamma)}\right)^\dagger\right]\\
&=&  \frac{1}{d^n} \sum_b \omega^{-[v,b]-\tilde{\Phi}^{(\gamma)}_g(b)} \,\left(T_{S_gb}^{(\gamma)}\right)^\dagger \\
&=&  \frac{1}{d^n} \sum_b \omega^{-[v+x_g,b]} \,\left(T_{S_gb}^{(\gamma)}\right)^\dagger \\
&=&  \frac{1}{d^n} \sum_b \omega^{-[S_gv + S_gx_g,b]} \,\left(T_b^{(\gamma)}\right)^\dagger \\
&=& A^{(\gamma)}_{S_gv + S_gx_g}
\end{array}
$$
This is Eq.~(\ref{CovarC3}) with $a_g=S_gx_g$.
$\Box$\medskip

Theorem~\ref{C1b-cov} has the following generalization, which we will refer to in Section~\ref{1vs2}.
\begin{Cor}\label{C1c-cov}
	For any number  $n\geq 1$ of qudits of local dimension $d\geq 2$, be $G \subseteq \text{Cl}_n$ a subgroup of the Clifford group such that ${\cal{P}}_n \subset G$. A $G$-covariant Wigner function according to \hyperlink{OB}{(OB)} exists if and only if  $[\Phi_{\text{\upshape cov}}] =  0 \in H^1(Q,U_{\text{\upshape cov}})$.
\end{Cor} 
The proof is exactly the same as for Theorem~\ref{C1b-cov}.

\subsection{Even vs. odd dimension}

For odd dimension $d$, and all numbers $n$ of qudits, a Clifford-covariant Wigner function satisfying the conditions \hyperlink{OB}{(OB)}, \hyperlink{SW1}{(SW1)}--\hyperlink{SW4}{(SW4)} has been explicitly constructed in \cite{GrossThesis}. In the present formalism, the existence of a Clifford-covariant Wigner function satisfying \hyperlink{OB}{(OB)} follows by Observation~\ref{PhiOdd} and Theorem~\ref{C1b-cov}.\medskip

In even dimension we have the following result.
\begin{Theorem}\label{WiCovEven}
If the local dimension $d$ is even, then, for any number $n$ of local systems, Clifford-covariant Wigner functions satisfying \hyperlink{OB}{(OB)} do not exist.
\end{Theorem}

{\em{Remark:}} There is a related no-go theorem in~\cite{Schmid}, addressing the impossibility of positive representation of the stabilizer sub-theory in even dimension. Here, we assume much less about Wigner functions than the related no-go theorem in~\cite{Schmid}, namely only (OB). We point out specifically that the assumption of diagram preservation is not required.

Furthermore, a special instance of Theorem~\ref{WiCovEven} was proved in \cite{Zhu2}, namely for the multi-qubit case of $n>1$, $d=2$. This result is based on the fact---proved in the same paper---that the multi-qubit Clifford groups are unitary 3-designs.  However, this property does not extend to other even dimension except powers of 2 \cite{ZachWebb}.
\smallskip

{\em{Remark:}} From the perspective of QCM, Theorem~\ref{WiCovEven} can be bypassed. Namely, as explained in Section~\ref{vari}, no computational power is lost in QCM if the Clifford gates are dispensed with. This holds even though such gates are by convention part of QCM \cite{BK,NegWi}. If no Clifford gates occur, then the breakdown of Clifford covariance is no problem for the classical simulation. The computational power, and hence the hardness of classical simulation, rests with the Pauli measurements.\medskip

To prove Theorem~\ref{WiCovEven}, we first establish the following statement about the cohomology class $[\Phi_{\text{\upshape cov}}]$ for the $n$-qudit Clifford group.

\begin{Lemma}\label{PhiEven}
	For all even local dimensions $d$ and all qudit numbers $n$, it holds that $[\Phi_{\text{\upshape cov}}]\neq 0$.
\end{Lemma}

\noindent{\em{Proof of Lemma~\ref{PhiEven}.}} The basic proof strategy is to identify a group element $g \in \text{Cl}_n$ and a 2-chain $f\in \tilde{C}_2$ such that
\begin{equation}\label{gfCond}
g\partial f = \partial f,\;\text{and} \; \tilde{\Phi}_g(\partial f) \neq 0.
\end{equation}
Assume Eq.~(\ref{gfCond}) holds and $[\Phi_\text{cov}]=0$. From the latter, $\tilde{\Phi}_g(\partial f) = \Phi_\text{cov}([g],\partial f) = \nu(\partial f) - \nu(g\partial f)$, for some $\nu \in C^0(G,U_\text{cov})$. Therein, with Eq.~(\ref{gfCond}), $0\neq 0$---Contradiction. Thus, the existence of a pair $g,f$ satisfying Eq.~(\ref{gfCond}) implies that $[\Phi_\text{cov}]\neq 0$.

In the following, we show that $g$, $f$ can be chosen in accordance with Eq.~(\ref{gfCond}), for all even $d$. We focus on $n=1$; replacing $g$ by $g\otimes I^{\otimes n-1}$ and all $a_z,a_x\in\mathbb{Z}_d$ by $(a_z,0,\ldots,0),(a_x,0,\ldots,0)\in\mathbb{Z}_d^n$ immediately generalizes the proof to all $n\in\mathbb{N}$. For concreteness, we set $\gamma(a)=a_za_x$, but our final statement is independent of the phase convention $\gamma$.

For $d=4m+2$, $m\in\mathbb{N}_0$, we consider the Fourier transform $g=\frac{1}{\sqrt{d}} \sum_{k,l=0}^{d-1} \omega^{kl} |k\rangle \langle l|$, which is a Clifford unitary that acts by conjugation as
\begin{equation}
	g\,\cdot\, g^\dagger: X \longrightarrow Z \longrightarrow X^{-1} \longrightarrow Z^{-1} \longrightarrow X.
\end{equation}
Further, $f:=[u|v]$, with $u=(u_z,u_x)$ and $v=(v_z,v_x)$ by $u_x=v_z=0$ and $u_z=v_x=2m+1$. Then
\begin{equation}
\begin{alignedat}{5}
	T_u&=Z^{2m+1},& T_v&=X^{2m+1} &T_{u+v}&=\sqrt{\omega}^{(2m+1)^2}Z^{2m+1}X^{2m+1},\\
	gT_ug^\dagger&=T_v, & \quad gT_vg^\dagger&=T_u,& \quad gT_{u+v}g^\dagger&=\omega^{-(2m+1)}T_{u+v}.
\end{alignedat}
\end{equation}
Hence, $g\partial[u|v]=\partial[u|v]$ but 
$\tilde{\Phi}_g(\partial[u|v])=0+0+(2m+1) = d/2  \neq 0\in \mathbb{Z}_d$. Thus, Eq.~(\ref{gfCond}) applies.\smallskip

For $d=4m$, $m\in\mathbb{N}$, we introduce the unitary $g=\frac{1}{d}\sum_{k,l=0}^{d-1}\omega^{mk^2}\left(\sum_{j=0}^{d-1}\omega^{mj^2+(k-l)j}\right)|k\rangle\langle l|$, which is in the Clifford group since
\begin{equation}
	gZg^\dagger=\sqrt{\omega}^{2m}ZX^{2m},\quad gXg^\dagger=\sqrt{\omega}^{\,-2m}Z^{2m}X.
\end{equation}
Choosing $f:=[u|v]$, with $u_z=v_z=1$, $u_x=0$, and $v_x=2m$ yields
\begin{equation}
\begin{alignedat}{5}
	T_u&=Z,& T_v&=\sqrt{\omega}^{2m}ZX^{2m} &T_{u+v}&=\sqrt{\omega}^{d}Z^2X^{2m},\\
	gT_ug^\dagger&=T_v, & \quad gT_vg^\dagger&=\omega^{2m}T_u,& \quad gT_{u+v}g^\dagger&=T_{u+v}.
\end{alignedat}
\end{equation}
Again, $\partial f = g\partial f$, and
$\tilde{\Phi}_g(\partial[u|v])=0+2m-0 = d/2  \neq 0\in \mathbb{Z}_d$.
Eq.~(\ref{gfCond}) thus applies. 
\hfill $\Box$\medskip

{\em{Proof of Theorem~\ref{WiCovEven}.}} The statement is the combined conclusion of Lemma~\ref{PhiEven} and Theorem~\ref{C1b-cov}. $\Box$

\subsection{First vs. second cohomology group}\label{1vs2}

The Clifford group splits when $d$ is odd and $n\geq 1$, or $d=2$ and $n=1$. It does not split for $d$ even and $n\geq 2$.
 \cite{appleby2005symmetric,GrossThesis,bolt1961cliffordII}. Splitting is also a notion of group cohomology, living at the second level. {\em{Is there a connection with Clifford covariance?}}

In this section we clarify that such a connection does indeed exist. However, it is not one-to-one. Namely, splitting of the symmetry group $G$ is necessary for $G$-covariance, but not sufficient. 

A notion of particular interest for this discussion is the specific faithful group action of the group $G$ (the Clifford group, or a subgroup thereof) on the phase space $V$ required by Definition~\ref{CC}. The existence of such a group action is a precondition for covariance. We establish that splitting is necessary and sufficient for the existence of the faithful group action. 

Recall that splitting is equivalent to $[\zeta]=0$ with $\zeta$ defined by Eq.~\eqref{MuPropC}, and $Q$ and $U_\text{cov}$ are defined above Eq.~(\ref{eq:Phi}). In summary we arrive at the following picture:
\begin{equation}\label{CohoRels}
\fbox{$
\begin{array}{ccc}\\
G\text{-covariance} & \Rightarrow & \parbox{3cm}{$\exists$ faithful group\\action of $G$ on $V$}\vspace{2mm}\\
\Updownarrow & & \Updownarrow \vspace{2mm}\\
{[\Phi_{\text{\upshape cov}}]} =  0 \in H^1(Q,U_{\text{\upshape cov}}) & \Rightarrow & [\zeta]=0 \in H^2(Q,E)\\ \mbox{ }
\end{array}
$}
\end{equation} 
The vertical arrow on the left is the content of Corollary~\ref{C1c-cov} in Section~\ref{CliffEx}. The remaining three arrows are established in Sections~\ref{Split} and \ref{crs} below.

\subsubsection{Splitting and faithful group action}\label{Split}

With $\zeta$ and $Q$ as defined in Section \ref{sec:splitting-group-cocycle}, we have the following result.

\begin{Theorem}\label{fga}
For any given number $n$ of $d$-level systems, be $G \subseteq \operatorname{Cl}_n$ a subgroup of the Clifford group such that ${\cal{P}}_n \subset G$. A faithful action of $G$ on the phase space $V$ of the form $\tau:G\times V\rightarrow V$, $\tau_g(v) = S_g v + a_g$ exists if and only if $[\zeta]=0\in H^2(Q,E)$.
\end{Theorem}

{\em{Proof of Theorem~\ref{fga}.}} ``Only if'': Assume that a faithful group action of the form $\tau_g(v) = S_g v + a_g$ does exist.

First, we examine $\tau_g$ for $g\in{\cal{P}}_n=\{t_a,a\in E\}$, where $t_a$ denotes $T_a$ with the phase modded out. For readability, we write $y(a)$ instead of $a_g$ with $g=t_a$. Note that $t_at_b=t_{a+b}$ and $t_a^\dagger=t_{-a}$. Therefore, $\forall g\in\{t_a\}$, $S_g$ as defined in Eq.~\eqref{eq:symmetry} is the identity transformation, i.\,e., 
$\tau_{t_a}(v)=v+y(a)$.

Using that $\tau_g$ is a group action, we find that $y: E\cong \mathbb{Z}^{2d}\rightarrow V\cong \mathbb{Z}^{2d}$ is linear:
$$
	y(a+b)=\tau_{t_at_b}(v)-v=\tau_{t_a}(\tau_{t_b}(v))-v = y(a)+y(b)\;\;\forall a,b\in E
$$
Moreover, $y$ is injective and, thus, invertible:
$$
	y(a)=y(b)\;\;\Rightarrow\;\; \tau_{t_a}=\tau_{t_b}\;\;\Rightarrow\;\; t_a=t_b \;\;\Leftrightarrow\;\; a=b,
$$
where the second step relies on $\tau$ being faithful.

Thanks to the invertibility of $y$, we can now choose a section $\theta:Q\rightarrow G$, see Eq.~\eqref{MuProp}, such that $\tau_{\theta(S)}(v)=Sv$. In general, $\tau_{\theta'(S)}(v)=Sv+ a_{\theta'(S)}$. However, by Eq.~\eqref{muCha}, we can always transition from $\theta'(S)$ to $\theta(S)=t_{y^{-1}[-a_{\theta'(S)}]}\theta'(S)$. This ensures that
$\tau_{\theta(S)}(v)=\tau_{t_{y^{-1}[-a_{\theta'(S)}]}}(\tau_{\theta'(S)}(v))=Sv$, 
as desired. 
By fixing $\theta$ we single out a specific representative $\zeta$ from the equivalence class $[\zeta]\in H^2(Q,E)$.

At last, we show that, for this choice of $\theta$, $\zeta(S_1,S_2)=0$ $\forall S_1,S_2\in Q$. Using Eq.~\eqref{MuProp} to go from the first to the second line, we find
$$
\begin{array}{rcl}
	S_1S_2 v &=& \tau_{\theta(S_1)}(\tau_{\theta(S_2)}(v)) =\tau_{\theta(S_1)\theta(S_2)}(v)\\
	&=&\tau_{t_{\zeta(S_1,S_2)}\theta(S_1S_2)}(v)=\tau_{t_{\zeta(S_1,S_2)}}(\tau_{\theta(S_1S_2)}(v))=S_1S_2v+y(\zeta(S_1,S_2)), \;\;\forall S_1,S_2\in Q.
\end{array}
$$
Hence $y(\zeta(S_1,S_2))=0$ and, since $y$ is linear and invertible, $\zeta(S_1,S_2)=0$ $\forall S_1,S_2 \in Q$. This, finally, yields $[\zeta]=0\in H^2(Q,E)$. 

``If'': Assume that $[\zeta]=0\in H^2(Q,E)$. Then we can choose $\theta$ such that $\zeta(S_1,S_2)=0$ $\forall S_1,S_2\in Q$. According to Eq.~\eqref{MuProp}, any $g\in G$ can be cast in the form $g=t_{\alpha_g}\theta(S_g)$. 

Let us show that $\tau_g(v)=S_g v + \alpha_g$ is a faithful group action of $G$ on $V$. Observing that $hg=t_{\alpha_h}\theta(S_h)t_{\alpha_g}\theta(S_g)=t_{\alpha_h}\theta(S_h)t_{\alpha_g}(\theta(S_h))^\dagger\theta(S_h)\theta(S_g)=t_{\alpha_h+S_h\alpha_g}\theta(S_hS_g)$,
we find
$\tau_{hg}(v)=\tau_{t_{\alpha_h+S_h\alpha_g}\theta(S_hS_g)}(v)=S_hS_g v + \alpha_h+S_h\alpha_g =S_h(S_gv+\alpha_g)+\alpha_h= \tau_h(\tau_g(v))$,
i.\,e., $\tau$ is a group action.

Moreover, $\tau$ is faithful. Let $\tau_g(v)=\tau_h(v)$, $\forall v\in V$. Then $\alpha_g=\alpha_h$, $S_g=S_h$, and, thus, $g=t_{\alpha_g}\theta(S_g)=t_{\alpha_h}\theta(S_h)=h$. 
\hfill$\Box$\smallskip

This establishes the vertical arrow on the right of the diagram Eq.~(\ref{CohoRels}).

\subsubsection{Covariance implies splitting} \label{crs}

We have the following result, as a corollary to Theorem~\ref{fga}:
\begin{Cor}\label{C1d-cov}
	For any given number $n$ of $d$-level systems, be $G \subseteq \text{Cl}_n$ a subgroup of the Clifford group such that ${\cal{P}}_n \subset G$. A $G$-covariant Wigner function satisfying \hyperlink{OB}{(OB)} exists only if $[\zeta]=0\in H^2(Q,E)$.
\end{Cor}

{\em{Proof of Corollary~\ref{C1d-cov}.}} We first show that $G$-covariance as defined in Definition~\ref{CC} implies the existence of a faithful group action of $G$ on $V$ of the form $\tau:G\times V\rightarrow V$, $\tau_g(v) = S_g v + a_g$. 

A Wigner function satisfying \hyperlink{OB}{(OB)} is Clifford covariant if and only if $g(A_v)=A_{S_gv+a_g}=A_{\tau_g(v)}$ $\forall v\in V$, $\forall g\in G$, cf. Eq.~\eqref{CovarC3}. Therefore
\begin{equation}
	A_{\tau_{hg}(v)}=(hg)(A_v)=h(g(A_v))=A_{\tau_h(\tau_g(v))}\;\;\forall v\in V,\;\forall g,h\in G.
\end{equation} 
Since the $A_v$ constitute an operator basis, $A_v=A_{v'}$ implies $v= v'$. Thus, $\tau_{hg}(v)=\tau_h(\tau_g(v))$ $\forall v\in V$, $\forall g,h\in G$, and $\tau$ is, indeed, a group action. 

If $\tau$ were not faithful, there would exist $g,h\in G$ with $g\neq h$ such that
\begin{equation}
	g(A_v)=h(A_v)\;\;\forall v\in V \;\;\Leftrightarrow\;\;h^{-1} gA_v = A_vh^{-1}g \;\;\forall v\in V.
\end{equation}
Using, again, that $\{A_v,v\in V\}$ is a basis, we observe that $[h^{-1} g, Y]=0$ for any linear operator $Y$ and, thus, the unitary $h^{-1}g$ is, up to phase, the identity operator. This contradicts that $g\neq h$ in~$G$. Hence, $\tau$ is a faithful group action.

With Theorem~\ref{fga}, it follows that $[\zeta]=0$. \hfill$\Box$\medskip

Corollary~\ref{C1d-cov} demonstrates the upper horizontal arrow in the diagram Eq.~(\ref{CohoRels}). In sum, we have established the two vertical arrows and the upper horizontal one, implying the remaining lower horizontal arrow. This completes the diagram Eq.~(\ref{CohoRels}).\medskip

The lower horizontal arrow, so far derived through reasoning about the physics concept of Wigner function, relates the cohomology groups $H^1(Q,U_{\text{\upshape cov}})$ and $H^2(Q,E)$, which are purely mathematical objects. In Appendix~\ref{Coho12} we independently establish this relation as a mathematical fact, without recurring to physics concepts.

\subsubsection{Splitting does not imply covariance} 

Splitting of a Clifford subgroup $G$ does not guarantee the existence of a $G$-covariant Wigner function. 
This is demonstrated with the following 1-qubit example: We consider the covariance group $G$ generated by the Hadamard gate $H$ and the one-qubit Pauli group. This group splits, $G=\langle H\rangle \ltimes P_1$.  However, $[\Phi_\text{cov}]\neq 0$. Namely, consider the Pauli operators $T_{a_1}=X$, $T_{a_2}=Y$, $T_{a_3}=Z$. Then, $\tilde{\Phi}_H(a_1)=\tilde{\Phi}_H(a_3)=0$, and $\tilde{\Phi}_H(a_2)=1$. Thus, $\Phi_\text{cov}(H,\partial [a_1|a_3]) = \tilde{\Phi}_H(\partial [a_1|a_3])=1$ and $H \partial [a_1|a_3] = \partial [a_1|a_3]$. These relations imply that $[\Phi_\text{cov}]\neq 0$ (see the proof of Lemma~\ref{PhiEven}). Thus, with Corollary~\ref{C1c-cov}, no $G$-covariant Wigner function satisfying (OB) exists.

\section{Positive representation of Pauli measurement}\label{PRPM}

In this section we first define ``positive representation of Pauli measurement'', and then establish a necessary and sufficient condition for it, applicable to all Wigner functions that satisfy \hyperlink{OB}{(OB)}, \hyperlink{SW1}{(SW1)}--\hyperlink{SW4}{(SW4)}. Finally, we apply this result to even and odd dimensions, making use of the respective structure theorems for Pauli observables.

\subsection{When are Pauli measurements positively represented?}\label{PRPM1}

Denote by $\Pi_{a,s}$ the projector associated with the outcome $s\in \mathbb{Z}_d$ in the measurement of the Pauli observable $T_a$, i.\,e., $\Pi_{a,s}$ is the projector onto the eigenspace of $T_a$ with eigenvalue $\omega^s$. The probability to obtain $s$ when measuring $T_a$ on the state $\rho$ is $\text{Tr}(\Pi_{a,s} \rho)$. We then have the following definition of ``positive representation of Pauli measurement'':

\begin{Def}\label{PosR} A pair $(W,\Theta)$ of a Wigner function $W$ and an effect function $\Theta$ satisfying \hyperlink{SW4}{(SW4)}, i.\,e., $
\text{\upshape{Tr}}(\Pi_{a,s}\rho) = \sum_{v\in V} \Theta_{\Pi_{a,s}}(v) W_\rho(v),
$ represents Pauli measurements positively if the following two properties hold.
 \begin{itemize}
 \item[(a)]{For all Pauli measurements, the corresponding effect functions $\Theta_{\Pi{a,s}}: V\longrightarrow \mathbb{R}$ satisfy
 \begin{equation}\label{Theta2} 
 \Theta_{\Pi_{a,s}}(v)\geq 0, \; \forall v\in V,\,\forall a,s.
 \end{equation} }
 \item[(b)]{For all Pauli measurements, the non-negativity of the Wigner function is preserved under measurement, i.e.,
 \begin{equation}\label{PosPres}
 W_\rho \geq 0 \Longrightarrow W_{\Pi_{a,s}\rho \Pi_{a,s}}\geq 0, \forall a,s.
 \end{equation} }
 \end{itemize}
\end{Def}
The above definition of ``positive representation of Pauli measurement'' is intuitive. Condition (a) says that the effects $\Theta_{\Pi_{a,s}}$ associated with all Pauli measurements are non-negative, and if the Wigner function $W_\rho$ is non-negative as well, then the outcome probabilities for Pauli measurements can be obtained by sampling from the phase space $V$. Condition (b) says that if a state $\rho$ is represented by a non-negative Wigner function, then for any Pauli measurement with any outcome, the post-measurement state is also represented by a non-negative Wigner function.

\subsection{Cohomological condition for positive representation}\label{sec:CohCondPositivity}

Here we show the following.

\begin{Theorem}\label{M_PosRep}
For any system of $n$ qudits with $d$ levels, $n,d\in \mathbb{N}$, a pair of Wigner and effect function satisfying \hyperlink{OB}{(OB)}, \hyperlink{SW1}{(SW1)}--\hyperlink{SW4}{(SW4)} that represents Pauli measurement positively exists if and only if $[\beta]=0 \in H^2(\cC,\mathbb{Z}_d)$.
\end{Theorem}
The proof of Theorem~\ref{M_PosRep} proceeds in several steps. To begin, we note that $(T_b)^k\sim T_{bk}$, and define phases $\varphi_b:\mathbb{Z}_d \longrightarrow \mathbb{Z}_d$, for all $b$, such that
\begin{equation}\label{VarphiDef}
(T_b)^k = \omega^{\varphi_b(k)}T_{bk},\;\; \forall b,\;\forall k.
\end{equation}
With Ansatz~\eqref{PhaPoDef}, Reality \hyperlink{SW1}{(SW1)} becomes equivalent to $c_b^* T_b = c_{-b}T_{-b}^\dagger$, and hence to both
\begin{equation}
\label{Herm}
c_b^* =  \omega^{\varphi_{-b}(-1)}c_{-b}\;\;\text{and}\;\;c_b^* =  \omega^{\varphi_{b}(-1)}c_{-b}.
\end{equation}
To prepare for subsequent applications, we 
observe that $(T_b)^{kl} = \left((T_b)^k\right)^l$, for all $k,l\in \mathbb{Z}_d$ and all $b$. For the phases $\varphi_b$ defined in Eq.~(\ref{VarphiDef}) this entails
$$
\varphi_b(kl) = \varphi_{kb}(l) + l\varphi_b(k).
$$
We will later make use of this relation for the special case of $l = -1 \mod d$,
\begin{equation}\label{Asso}
\varphi_b(-k) = \varphi_{kb}(-1) - \varphi_b(k).
\end{equation}

As the next step towards proving Theorem~\ref{M_PosRep}, we have the following two results.
\begin{Lemma}\label{Real}
	For all pairs of Wigner and effect functions satisfying \hyperlink{OB}{(OB)}, \hyperlink{SW3}{(SW3)}, \hyperlink{SW4}{(SW4)}, the functions $\Theta_{\Pi_{a,s}}(\cdot)$ are, for all $a,s$, of the form
	\begin{equation}\label{Theta_2}
		\Theta_{\Pi_{a,s}}(v) = \frac{1}{d} \sum_k  \omega^{-ks'}c_{a}(k)
	\end{equation}
	with $s':=s+[v,a]$, $c_a(k):=\omega^{\varphi_a(k)}c_{ka}$, and
	\begin{equation}\label{Herm2}
		c_a(k)^* =  c_a(-k).
	\end{equation}
	In particular, all $\Theta_{\Pi_{a,s}}(\cdot)$ are real-valued.	
\end{Lemma}
Further,
\begin{Lemma}\label{Mag1}
For any system of $n$ qudits with $d$ levels, $n,d\in \mathbb{N}$, a pair of Wigner and effect function satisfying \hyperlink{OB}{(OB)}, \hyperlink{SW1}{(SW1)}--\hyperlink{SW4}{(SW4)} can represent measurement positively only if $|c_b|=1$, $\forall b$.
\end{Lemma}
{\em{Proof of Lemma~\ref{Real}.}} For any $a$, the projector $\Pi_{a,s}$ can be represented as
$$
\Pi_{a,s} = \frac{1}{d} \sum_{i=0}^{d-1} \omega^{-si}(T_a)^i.
$$
Starting from the expression for $\Theta$ in Eq.~(\ref{ThetaTrace}) and the expansion of $A_v$ in Eq.~(\ref{PhaPoDef}), which invoke the assumptions  \hyperlink{OB}{(OB)}, \hyperlink{SW3}{(SW3)}, and \hyperlink{SW4}{(SW4) of the lemma}, we obtain by direct computation 
\begin{equation}\label{Theta_1}
\Theta_{\Pi_{a,s}}(v) = \frac{1}{d} \sum_{k=0}^{d-1} \omega^{-k(s+[v,a])+\varphi_a(k)}c_{ka}.
\end{equation}
We have thus established all $\Theta_{\Pi_{a,s}}$ as functions from $V$ to $\mathbb{C}$. We still need to show that all $\Theta_{\Pi_{a,s}}(v)$ are real-valued. To this end, we use the above definitions $s'=s +[v,a]$ and $c_a(k)=\omega^{\varphi_a(k)}c_{ka}$, which simplify $\Theta_{\Pi_{a,s}}(v)$ to Eq.~(\ref{Theta_2}).

The $\Theta_{\Pi_{a,s}}(v)$ are thus real-valued if $c_a(k)^* = c_a(-k)$, for all $a$ and all $k$. This property  we now demonstrate.
$$
\begin{array}{rcl}
c_a(k)^* &=& c_{ka}^* \, \omega^{-\varphi_a(k)}\\
&=& c_{-ka}  \, \omega^{\varphi_{ka}(-1)-\varphi_a(k)}\\
&=&  c_a(-k) \, \omega^{-\varphi_a(-k) + \varphi_{ka}(-1)-\varphi_a(k)}\\
&=& c_a(-k)
\end{array}
$$
Therein, the first and fourth line follow by the definition of $c_a(\cdot)$, and the second line by Eq.~(\ref{Herm}). Finally, using Eq.~(\ref{Asso}) in the last relation, we obtain Eq.~(\ref{Herm2}). $\Box$
\medskip

{\em{Proof of Lemma~\ref{Mag1}.}} The proof consists of two parts. Part (a) invokes the condition Eq.~(\ref{Theta2}) in Def.~\ref{PosR}, and leads to Eq.~(\ref{c_aCond2}). Part (b) invokes the condition Eq.~(\ref{PosPres}) in Def.~\ref{PosR} of positive representation of Pauli measurement, and leads to Eq.~(\ref{c_aCond1}) below.  Both constraints together imply the statement of the lemma.

{\em{(a).}} First we exclude the possibility of $|c_a|>1$ for any $a$. With the expansion of $A_v$ in Eq.~(\ref{PhaPoDef}), for any $v\in V$,
$|c_a| = \left| \text{Tr}(T_a A_v) \right| = \left| \sum_{s} \omega^s \text{Tr}(\Pi_{a,s} A_v) \right| \leq \sum_{s} \left|\text{Tr}(\Pi_{a,s} A_v)\right| = \sum_{s} \Theta_{a,s}(v) = 1$. 
Herein, the third relation follows by the triangle inequality, and $\left|\text{Tr}(\Pi_{a,s} A_v)\right| = \text{Tr}(\Pi_{a,s} A_v)=\Theta_{\Pi_{a,s}}(v)$ relies on Eq.~\eqref{ThetaTrace} and the assumption of positive representation; cf. Eq.~(\ref{Theta2}) in Def.~\ref{PosR}. The last equality is by Eq.~(\ref{ThetaSum}). Thus we arrive at
\begin{equation}\label{c_aCond2}
|c_a| \leq 1,\;\; \forall a.
\end{equation}
{\em{(b).}} Consider the expectation value $\langle T_a \rangle_\rho$. With Eqs.~(\ref{OB}) and (\ref{PhaPoDef}),
$$
\langle T_a\rangle_\rho = \sum_{v\in V} W_\rho(v) \omega^{-[v,a]} c_a.
$$
Now assume that $|c_a|<1$, and consider quantum states $\rho$ that are positively represented by $W$, i.e., $W_\rho \geq 0$. Then,
$$
\begin{array}{rcl}
|\langle T_a\rangle_\rho | &\leq& \sum_{v\in V} |W_\rho(v)| |c_a|\\
 &<& \sum_{v\in V} |W_\rho(v)|\\
 &=& \sum_{v\in V} W_\rho(v)\\
 &=& 1.
\end{array}
$$
In short,  $|\langle T_a\rangle_\rho | < 1$, for all $a\in E$. Above, the first line holds by the triangle inequality, the second line invokes the assumption $|c_a|< 1$, the third line invokes the other assumption $W_\rho\geq 0$, and the fourth line follows from the Standardization condition \hyperlink{SW2}{(SW2)}. 

In view of this constraint, consider any eigenstate $\rho(a,s)$ of the Pauli observable $T_a$, with an eigenvalue $\omega^s$. Hence, $|\langle T_a\rangle_{\rho(a,s)} | =1$.  With the above, $|c_a|<1$ implies that $W_{\rho(a,s)}<0$, $\forall\, s\in \mathbb{Z}_d$. 

With Eqs.~\eqref{OB}-\eqref{StandBas}, we observe that the completely mixed state $I/d^n$ has a positive Wigner function. However, the post-measurement states resulting from measuring $T_a$ on $I/d^n$ are of the type $\rho(a,s)$. Thus, if $|c_a|<1$ then $W$ is not positivity preserving under measurement of the Pauli observable $T_a$, contradicting the condition Eq.~(\ref{PosPres}) in Def.~\ref{PosR}. Hence, a Wigner function represents Pauli measurement positively only if 
\begin{equation}\label{c_aCond1}
|c_a|\geq 1,\;\; \forall a.
\end{equation}

Combining Eqs.~(\ref{c_aCond1}) and (\ref{c_aCond2}) leaves $|c_a|=1$ as the only option, for all $a$. $\Box$\medskip

As the final step in proving Theorem~\ref{M_PosRep}, we need to constrain the phases of the coefficients $c_a(k)$, cf. Lemma~\ref{Real}. This requires a discrete version of Bochner's theorem.

\begin{Lemma}\label{Bochner}(Variation on Bochner's theorem) For a given function $f:\mathbb{Z}_d \longrightarrow \mathbb{C}$ the Fourier transform $\hat{f}$ of $f$ is non-negative if and only if the matrix $M$ with coefficients
\begin{equation}\label{DefM}
M^x_y = f(x-y),\;\; \forall x,y \in \mathbb{Z}_d,
\end{equation}
is positive semidefinite.
\end{Lemma}
The proof of Lemma~\ref{Bochner} is the same as in \cite{GrossThesis} (Theorem 44 therein). To explicitly demonstrate that it applies in both odd and even dimensions, we restate it in Appendix~\ref{Boch}. 

We are now ready to prove the main result of this section, Theorem~\ref{M_PosRep}.\smallskip

{\em{Proof of Theorem~\ref{M_PosRep}.}} ``Only if'': We assume that a given pair of Wigner function $W$ and effect function $\Theta$ satisfying \hyperlink{OB}{(OB)}, \hyperlink{SW1}{(SW1)}--\hyperlink{SW4}{(SW4)} represents Pauli measurement positively. Our first goal is to show that the coefficients $c_a(k)$ may then be expressed in the form
\begin{equation}\label{cak}
c_a(k) = \left(\omega^{r_a}\right)^k,\; \text{with}\; r_a\in \mathbb{Z}_d,\;\forall a.
\end{equation}
The set $\{r_a, \forall a\}$ characterizes the phase point operators $\{A_v\}$ via $c_a=c_a(1)=\omega^{r_a}$. 

With Lemma~\ref{Mag1}, we can express the coefficients $c_a(k)$ as
$c_a(k) = e^{i\chi_a(k)},\; \chi_a(k) \in \mathbb{R}$. We furthermore observe that, with Eqs.~(\ref{Stand}) and~(\ref{VarphiDef}), it holds that $c_a(0)=1$, $\chi_a(0)=0$, for all $a \in E$. If $d=2$, then Eq.~(\ref{cak}) follows directly from Eq.~(\ref{Herm2}); namely the $c_a(k)$ are all real. 

For $d>2$, we consider the submatrix $M(a)|_{J\times J}$ for the set of rows (and columns) $J=\{1,2,k+1\}$, which, using Eq.~(\ref{Herm2}), reads
$$
M(a)|_{J\times J} = 
\left( \begin{array}{ccc}   
1 & e^{i\chi_a(1)} & e^{i\chi_a(k)}\\
e^{-i\chi_a(1)} & 1 & e^{i\chi_a(k-1)}\\
e^{-i\chi_a(k)} & e^{-i\chi_a(k-1)} & 1
\end{array}\right).
$$
By Lemmas~\ref{Real} and~\ref{Bochner}, we require the determinant of this matrix to be non-negative, which leads to the constraint
$$
e^{i\left(\chi_a(1)+\chi_a(k-1)-\chi_a(k)\right)} + \text{c.c.} \geq 2.
$$ 
The only solution of that constraint is $\chi_a(k)=\chi_a(k-1) +\chi_a(1)$, which we may use as a recursion relation for computing the angles $\chi_a(k)$, for all $k$. With $\chi_a(0)=0$,
$$
\chi_a(k) = k\, \chi_a(1).
$$
With the relation $c_a(k)^*=c_a(-k \mod d)$ we further find
$$
d \chi_a(1) = 0 \mod 2\pi,\;\; \forall a.
$$ 
Eq.~(\ref{cak}) follows from the last two relations, where $\omega^{r_a}=e^{i\chi_a(1)}$. With Eq.~(\ref{Theta_2}) this implies
\begin{equation}\label{ThetaExpl}
\Theta_{\Pi_{a,s}}(v) = \delta_{r_a,s+[v,a]}.
\end{equation}
Now consider the simultaneous measurement of the commuting observables $T_a$, $T_b$ with outcomes $s(a)$, $s(b)$ on the completely mixed state. We denote the resulting state by $\rho$. A further measurement of $T_a$ or $T_b$ on $\rho$ must produce outcomes $s(a)$, $s(b)$ with certainty, and a measurement of $T_{a+b}$ must produce the outcome $s_{a+b}=s(a)+s(b)-\beta(a,b)$ with certainty. Since the completely mixed state is positively represented and, by assumption, Pauli measurements are positivity preserving, it holds that $W_{\rho}\geq 0$. Using Standardization \hyperlink{SW2}{(SW2)}, we observe that for
any phase space point $v$ in the support of $W_{\rho}$, of which there is at least one, it must hold that $\Theta_{\Pi_{a,s(a)}}(v) = \Theta_{\Pi_{b,s(b)}}(v)=\Theta_{\Pi_{a+b,s(a)+s(b)-\beta(a,b)}}(v)=1$.  Therefore,  with Eq.~(\ref{ThetaExpl}),
$$
\begin{array}{rcl}
s(a) &=& r_a - [v,a],\\
s(b) &= &r_b - [v,b],\\
s(a)+s(b)-\beta(a,b) & =& r_{a+b} - [v,a+b].
\end{array}
$$
Adding the first two equations and subtracting the third, we obtain
\begin{equation}\label{rConstr}
r_a + r_b -r_{a+b} =\beta(a,b);\;\; \forall a,b \;\;\text{with}\;[a,b]=0.
\end{equation}
Note that in this condition the dependence on the particular phase space point $v$ has disappeared. In cohomological notation, Eq.~(\ref{rConstr}) reads $\beta = dr$; hence $[\beta]=0$.\medskip

``If'': Assume that $[\beta]=0$ holds. We show that then Gross' Wigner function, see Eq.~\eqref{eq:Gross}, has all the desired properties. We can choose a gauge such that $\beta\equiv 0$. In this gauge it holds that $\varphi_a \equiv 0$ for all $a$, cf. Eq.~(\ref{VarphiDef}). Note that $\varphi_a \equiv 0$, in turn, implies $(T_0)^0=T_0$ and, thus, $\mu^{\gamma(0)}=1$.
We assert
\begin{equation}\label{cSol}
c_{a} \equiv 1,
\end{equation}
and for this choice verify the Stratonovich-Weyl criteria as well as conditions Eq.~(\ref{Theta2}) and (\ref{PosPres}) of Def.~\ref{PosR}. 

(i) {\emph{Stratonovich-Weyl criteria.}} \hyperlink{OB}{(OB)}, \hyperlink{SW2}{(SW2)}, and \hyperlink{SW3}{(SW3)} hold since we use Ansatz~\eqref{PhaPoDef} and fulfill Conditions~\eqref{Stand} and~\eqref{Bas}. 
	With $\varphi_a\equiv 0$, \hyperlink{SW1}{(SW1)} in Ansatz~\eqref{PhaPoDef} becomes equivalent to $c_a^*=c_{-a}$, cf. Eq.~\eqref{Herm}, which is clearly satisfied by $c_a=1$ in Eq.~\eqref{cSol}. Finally, we ensure \hyperlink{SW4}{(SW4)} by setting $\Theta_Y(v)=\Tr(YA_v)$, cf. Eq.~\eqref{ThetaTrace}.

(ii) {\em{Condition Eq.~(\ref{Theta2}).}} With Eq.~(\ref{cSol}), we obtain $\Theta_{\Pi_{a,s}}=\delta_{s,[a,v]}$. Eq.~(\ref{Theta2}) is thus satisfied.

(iii) {\em{Condition Eq.~(\ref{PosPres}).}} It suffices to show that $\Pi_{a,s} A_v \Pi_{a,s} = \sum_{w\in V} q_v(w) A_w$, with all $q_v(w)$ real and non-negative. We have
\begin{equation}\label{ladder}
\begin{array}{rcl}
\Pi_{a,s} A_v \Pi_{a,s} &=& \displaystyle{\Pi_{a,s} \sum_{b|\, [a,b]=0} \frac{1}{d^n} \omega^{[b,v]} T_b^\dagger}\\
&=& \displaystyle{\left(\frac{1}{d} \sum_{k=0}^{d-1} \omega^{-ks}  (T_a)^k \right) \left(  \sum_{b|\, [a,b]=0} \frac{1}{d^n} \omega^{[b,v]} T_b^\dagger\right)}\\
&=&  \displaystyle{\frac{1}{d^{n+1}}  \sum_{k=0}^{d-1} \omega^{-ks} \sum_{b| [a,b]=0} \omega^{[b,v] } T_{b-ka}^\dagger }\\
&=&  \displaystyle{\frac{1}{d^{n+1}}  \sum_{k=0}^{d-1} \omega^{k([a,v]-s)} \sum_{b| [a,b]=0} \omega^{[b,v] } T_{b}^\dagger}\\
&=&  \displaystyle{\frac{\delta_{s,[a,v]}}{d^n}  \sum_{b} \delta_{[a,b],0} \, \omega^{[b,v] } T_{b}^\dagger}\\
&=&  \displaystyle{\frac{\delta_{s,[a,v]}}{d^{n+1}}  \sum_{k=0}^{d-1} \sum_{b} \omega^{[b,v]+k[b,a] } T_{b}^\dagger}\\
&=& \displaystyle{\delta_{s,[a,v]}\frac{1}{d}  \sum_{k=0}^{d-1} A_{v+ka}}\\
\end{array}
\end{equation}
Therein, in the first line we have used the ansatz Eq.~(\ref{cSol}). To remove the non-commuting elements ($[a,b]\neq 0$), we have used the following argument: $\Pi_{a,s} T_a =T_a\Pi_{a,s}= \omega^s \Pi_{a,s}$, and hence $\Pi_{a,s} T_b \Pi_{a,s}=\omega^{-s} \Pi_{a,s} T_a T_b \Pi_{a,s} = \omega^{-s+[a,b]}  \Pi_{a,s} T_b T_a \Pi_{a,s} = \omega^{[a,b]}  \Pi_{a,s} T_b \Pi_{a,s}$. Thus, if $[a,b]\neq 0$ then $\Pi_{a,s} T_b \Pi_{a,s}=0$. On the other hand, if $[a,b]=0$ then $\Pi_{a,s} T_b \Pi_{a,s} = \Pi_{a,s} \Pi_{a,s} T_b  = \Pi_{a,s} T_b $.
In the third line we have used the phase convention that yields $\beta \equiv 0$, and in the fourth line we re-organized the sum over $b$. Thus, all non-zero coefficients in the expansion of $\Pi_{a,s} A_v \Pi_{a,s}$ are positive.~$\Box$

\subsection{Even vs. odd dimension}\label{sec:PauliPosEvenOdd}

For odd local dimension, a Wigner function which represents Pauli measurement positively has been explicitly constructed \cite{GrossThesis}. In the framework established here, the existence of such a Wigner function follows by Observation~\ref{BetaOdd} and Theorem~\ref{M_PosRep}. 

For even local dimension, we have the following result.
\begin{Theorem}\label{PosReven}
For any system of $n\geq 2$ qudits with an even number $d$ of levels, a pair of Wigner and effect function satisfying \hyperlink{OB}{(OB)}, \hyperlink{SW1}{(SW1)}--\hyperlink{SW4}{(SW4)} that represents all Pauli measurements positively does not exist.
\end{Theorem}
To prove Theorem~\ref{PosReven}, we first establish the following fact. 
\begin{Lemma}\label{PauliThEven}
If the dimension $d$ is even, then for any number $n\geq 2$ of local systems it holds that $[\beta] \neq 0$.
\end{Lemma}
The proof of Lemma~\ref{PauliThEven} uses the existence of a Mermin square for all even $d$ and $n\geq 2$, yielding an alternate route to proving the non-existence of non-contextual ontological representations~\cite{Schmid,RaviThesis}.

\begin{table}
	\centering
	{\setlength{\tabcolsep}{3pt}
	\renewcommand{\arraystretch}{1.5}
	\begin{tabular}{rcl|c|rcl|c|rcl|cccr}
		\cellcolor{gray!15}$T_a$ &\cellcolor{gray!15}$=$& \cellcolor{gray!15}$Z^{-1} \otimes I$ & $\times$ & \cellcolor{gray!15}$T_b$& \cellcolor{gray!15}$=$ & \cellcolor{gray!15}$I\otimes Z$ & $\times$  & \cellcolor{gray!15}$T_{a+b}^{-1}$&\cellcolor{gray!15}$=$&\cellcolor{gray!15}$Z\otimes Z^{-1}$ & $=$ & $I$ & $\Rightarrow$ &$ \beta(a,b)=0$\\\hline
		&$\times $ &&&& $\times $ &&&& $\times$ &&&&&\\\hline
		\cellcolor{gray!15}$T_c$&\cellcolor{gray!15}$=$&\cellcolor{gray!15}$I\otimes \tilde{X}$ & $\times $ & \cellcolor{gray!15}$T_d$&\cellcolor{gray!15}$=$&\cellcolor{gray!15}$\tilde{X}^{-1}\otimes I$ & $\times$ & \cellcolor{gray!15}$T_{c+d}^{-1}$&\cellcolor{gray!15}$=$&\cellcolor{gray!15}$\tilde{X} \otimes \tilde{X}^{-1}$ &  $=$ & $I$ & $\Rightarrow$ & $\beta(c,d)=0$ \\\hline
		&$\times $ &&&& $\times $ &&&& $\times$ &&&&&\\\hline
		\cellcolor{gray!15}$T_{a+c}^{-1}$&\cellcolor{gray!15}$=$&\cellcolor{gray!15}$Z\otimes \tilde{X}^{-1}$ & $\times$ & \cellcolor{gray!15}$T_{b+d}^{-1}$&\cellcolor{gray!15}$=$&\cellcolor{gray!15}$\tilde{X} \otimes Z^{-1}$ & $\times$ & \cellcolor{gray!15}$T_{a+b+c+d}$&\cellcolor{gray!15}$=$&\cellcolor{gray!15}$\tilde{Y}^{-1}\otimes \tilde{Y}$ & $=$ & $I$ & $\Rightarrow$ & $\beta(a+c,b+d)=0$\\\hline
		& \veq &&&& \veq &&&& \veq &&&&&\\
		&$I$ &&&& $I$ &&&& $-I$ &&&&&\\
		& \vimpl &&&& \vimpl &&&& \vimpl &&&&&\\
		\multicolumn{3}{c|}{$\beta(a,c)=0$} && \multicolumn{3}{c|}{$\beta(b,d)=0$} && \multicolumn{3}{c|}{$\beta(a+b,c+d)=d/2$} &&&&
	\end{tabular}
	\renewcommand{\arraystretch}{1}}
	\caption{\label{GenSqTab}Mermin's square (shaded cells) generalized to arbitrary even dimension for the proof of Lemma~\ref{PauliThEven}. In each row and column of Mermin's square, the Pauli observables commute and imply a value of $\beta$ as stated. For the definition of $\tilde{X}$ and $\tilde{Y}$ see text.}
\end{table}

\begin{figure}[t]
	\centering
	\includegraphics[width=0.4\textwidth]{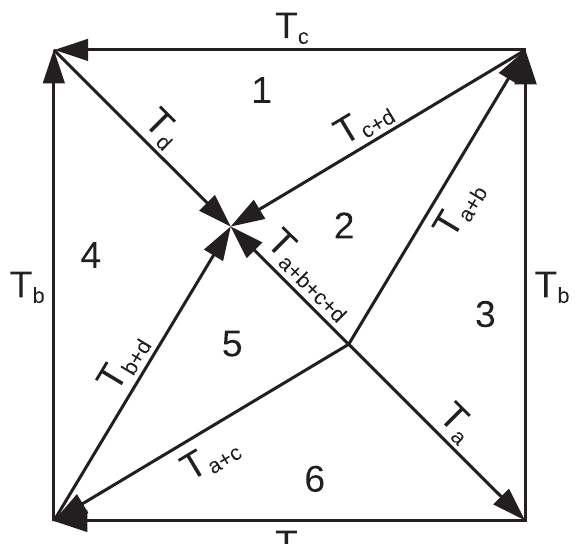}
	\caption{\label{GenSq}Topological reformulation of Mermin’s square: each row and column of Mermin's square corresponds to the boundary of an elementary face $f_j$ with $j\in\{1,\ldots,6\}$ as indicated. The exterior edges are identified as shown. The arrows give an orientation to the edges. For the explicit expressions for the Pauli observables appearing in this figure, see Table~\ref{GenSqTab}.}
\end{figure}

{\em{Proof of Lemma~\ref{PauliThEven}.}} The proof  proceeds by a construction generalizing Mermin's square to arbitrary even dimension. For any even $d$, define
$$
\tilde{X} = X^{d/2}, \tilde{Y} = {\sqrt{\omega}}^{d/2}\, X^{d/2} Z.
$$
Now consider the generalized Mermin square shown in Tab.~\ref{GenSqTab} and its equivalent topological reformulation depicted in Fig.~\ref{GenSq}.
The overall strategy of the proof is to identify a 2-cycle $F$ in the chain complex corresponding to Fig.~\ref{GenSq} such that $\partial F=0$ but $\beta(F)\neq 0$. Any such surface $F$ implies that $[\beta]\neq 0$. Namely, assume $[\beta]=0$, i.e., $\beta=d \gamma$ for some $\gamma \in C^1$. Then, for the above surface $F$, $\beta(F) = d \gamma(F) = \gamma(\partial F) = \gamma(0) = 0$, which contradicts the assumption $\beta(F)\neq 0$.

The overall surface $F\in C_2$ of the torus in Fig.~\ref{GenSq} has the required properties. More precisely, orienting $F$ such that all boundaries point counterclockwise and labeling the elementary faces $f_j$ as in Fig.~\ref{GenSq}, we observe that $F=\sum_{j=1}^6 f_j$ with $f_1=[c|d]$, $f_2=[a+b|c+d]$, $f_3=[a|b]$, $f_4=-[b|d]$, $f_5=-[a+c|b+d]$, and $f_6=-[a|c]$, and $\partial F=0$. It remains to be shown that $\beta(F)\neq 0$.
First, the top row of the square in Tab.~\ref{GenSqTab} reads $T_aT_bT_{a+b}^{-1}=I$; or equivalently, in the form matching Eq.~(\ref{3T}), $T_aT_b= \omega^0\,T_{a+b}$, such that $\beta(f_3)=0$. 
Now we turn to the rightmost column of Tab.~\ref{GenSqTab}, which is $T_{a+b}^{-1}T_{c+d}^{-1}T_{a+b+c+d}=-I$. Again we transform this into the normal form of Eq.~(\ref{3T}), which yields $\omega^{d/2}T_{a+b+c+d}=T_{a+b}T_{c+d}$ and $\beta(f_2) = d/2$. In the same fashion, we find $\beta(f_1)=\beta(f_4)=\beta(f_5)=\beta(f_6)=0$.
Indeed, $\partial F=0$ and $\beta(F) \mod d =d/2$. Hence $[\beta]\neq 0$.
 $\Box$\medskip

{\em{Proof of Theorem~\ref{PosReven}.}} The statement is the combined conclusion of Lemma~\ref{PauliThEven} and Theorem~\ref{M_PosRep}. $\Box$

\section{Discussion}\label{Disc}

Here we have addressed the following question about quantum computation with magic states (QCM). For qudits in odd dimension, negativity in the Wigner function of the initial magic state is a precondition for quantum speedup \cite{NegWi}.---Does the same hold in even dimension, e.g. for qubits?

This question has many facets. For example, it is known that if the notion of Wigner function is suitably generalized, namely to {\em{non-unique}} quasiprobability representations, then negativity in those representations can again be established as a precondition for quantum speedup \cite{RoM,QuWi19}.

In this paper, we have approached the question from a different angle. Namely, we  have investigated the more conventional Wigner functions derived from operator bases, a class to which the original Wigner function \cite{Wigner1932} and Gross' Wigner function in odd dimension \cite{GrossThesis} belong.  In the case of odd dimension relying on Gross' Wigner function (and likewise in \cite{RoM,QuWi19}), the key to establishing negativity as a precondition for quantum speedup are two structural properties of the quasiprobability representations involved---Clifford covariance and positive representation of Pauli measurement. Here we have shown that, in even dimension, no Wigner function constructed from operator bases has these important structural properties.   The obstructions to the existence of such Wigner functions are cohomological.

Specifically, the results of this paper are two-fold: First, for Wigner functions constructed from operator bases (and, in some cases, satisfying the Stratonovich-Weyl criteria in addition), we have formalized the obstructions to Clifford covariance and positive representation of Pauli measurement. These obstructions are cohomological in nature. Our most general result is Theorem~\ref{C1b-cov}, on obstructions to Clifford covariance. The only assumption made by the theorem is (OB), i.e., that the Wigner function in question is constructed from an operator basis. 

Theorem~\ref{M_PosRep} on the positive representation of Pauli measurement is the critical one from the perspective of QCM. Again, a cohomological obstruction is identified, but this time the theorem requires the Stratonovich-Weyl criteria as additional assumptions.

As the second set of our results, we have applied the above theorems to the case of even local dimension, where the cohomological obstructions don't vanish. Thereby we have extended the existing no-go results \cite{Zhu,Schmid,QuWi17} on the positive representation of the stabilizer sub-theory of quantum mechanics in even dimensions. The most general result so far is \cite{Schmid}, which shows that, under the assumption of diagram preservation, the stabilizer sub-theory cannot be positively represented. From this result, we have removed the assumption of diagram preservation. Specifically, in Theorem~\ref{WiCovEven} we have shown that, in all even dimensions, Clifford-covariant Wigner functions from operator bases do not exist. In Theorem~\ref{PosReven} we have shown that---whenever $n\geq 2$---Wigner functions from operator bases that also satisfy the Stratonovich-Weyl criteria cannot represent Pauli measurement positively. 
\smallskip

\begin{figure}
\begin{center}
\includegraphics[width=8cm]{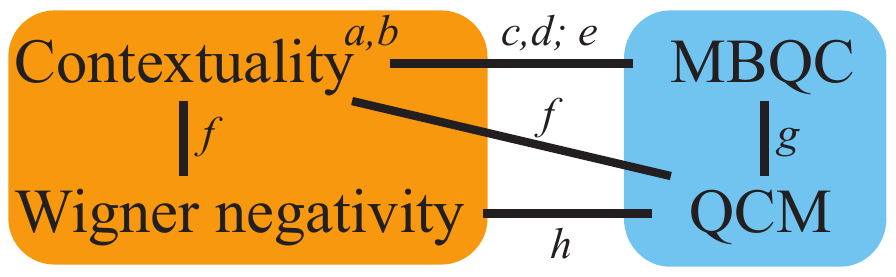}
\caption{\label{GraphSum}Canvas of cohomological properties in quantum computation with magic states (QCM) and measurement-based quantum computation (MBQC), relating to contextuality and Wigner function negativity. $a$=\cite{BMA}, $b$=\cite{Coho}: cohomological formulation of contextuality; $c$=\cite{AB}, $d$=\cite{RR13}: MBQC is contextual; $e$=\cite{CohoMBQC}: contextuality in temporally flat MBQCs is cohomological; $f$=\cite{Howard}: Wigner function negativity and state-dependent contextuality w.r.t. Pauli observables are the same in odd dimension; contextuality of the magic states is a precondition for quantum speedup; $g$=\cite{QuWi17}: computational power only resides in states and measurements, for both QCM and MBQC. $h$=\cite{NegWi}: in odd local dimension, quantum speedup in QCM requires Wigner function negativity. This result relies on the Clifford covariance of Gross' Wigner function, and its positive representation of Pauli measurement. [This work]: In even local dimension, cohomological invariants obstruct the existence of Wigner functions that are Clifford covariant and represent Pauli measurement positively.  }
\end{center}
\end{figure}

Taking a step back, we observed in the introduction that there is a web of cohomological facts connecting measurement-based quantum computation to contextuality\footnote{Any specific MBQC is contextual, i.e, cannot be described by a non-contextual hidden variable model, if it computes a non-linear Boolean function with sufficiently high probability of success \cite{AB,RR13}}, and beyond. This web of cohomological facts we have extended to a further computational scheme, quantum computation with magic states. Fig.~\ref{GraphSum} shows a map of these cohomological facts and their relations.

\paragraph{Acknowledgments.} RR and PF are funded from the Canada First Research Excellence Fund, Quantum Materials and Future Technologies Program. MZ is funded by the National Science and Engineering Research Council of Canada. CO is supported by the US Air Force Office of Scientific Research under award number FA9550-21-1-0002.

\bibliographystyle{quantum}
\bibliography{biblio}

\appendix

\section{Proof of Lemma~\ref{Bochner}}\label{Boch}

Lemma~\ref{Bochner} is a restatement of Theorem 44 in \cite{GrossThesis}. The entire chapter in \cite{GrossThesis} to which Theorem~44 belongs is written under the assumption that $d$ is odd. We restate the proof here, to clarify that for this particular lemma the assumption of odd $d$ is not needed.
\medskip

{\em{Proof of Lemma~\ref{Bochner}.}} Denote by $\nu_k$, $k \in \mathbb{Z}_d$, a character of $\mathbb{Z}_d$, $\nu_k(x) = \omega^{kx}$, for all $x \in \mathbb{Z}_d$; and by the same symbol the vector $\nu_k = (1,\omega^k,\omega^{2k},..,\omega^{(d-1)k})^T$. For any $d\times d$ matrix $M$ defined in Eq.~(\ref{DefM}) it holds that
$$
\begin{array}{rcl}
[M\nu_k]_y &=& \sum_xf(x-y) \omega^{kx}\\
 &=& \sum_xf(x) \omega^{kx} \omega^{ky}\\
 &=& d\, \hat{f}(k) \, \omega^{ky},
\end{array}
$$
and hence $\nu_k$ is an eigenvector of $M$ with eigenvalue $d\, \hat{f}(k)$. Since there are $d$ characters $\nu_k$, the matrix $M$ is diagonal in their basis. All its eigenvalues are non-negative if and only if $\hat{f}$ is non-negative. $\Box$

\section{Relation between the 1st and 2nd cohomology group of \texorpdfstring{$Q$}{Q}}\label{Coho12}

Here we provide an alternative derivation of the mathematical fact represented by the lower horizontal arrow in the diagram of Eq.~(\ref{CohoRels}), namely $[\Phi_\text{cov}] =0 \Longrightarrow [\zeta]=0$,  by purely mathematical reasoning that bypasses the discussion of Wigner functions. The result we obtain goes slightly beyond the one presented in the main text in that it applies to any symmetry group $G\subseteq \Cl_n$, regardless of whether $\mathcal{P}_n\subset G$. The idea is similar to the use of long exact sequences in cohomology that appears in Section 5.6 of \cite{Coho}.

We consider the exact sequence of $\ZZ_d$-modules
\begin{equation}\label{eq:exactsequence}
	0\to E \to  C^1 \to U_\cov \to 0,
\end{equation}
where the second map is defined by sending $a\in E$ to the $1$-cochain $[a,-]:E\to \ZZ_d$. There is an associated long exact sequence in cohomology
$$
\cdots H^1(Q,E) \to H^1(Q,C^1) \to H^1(Q,U_\cov) \xrightarrow{\sigma} H^2(Q,E) \to \cdots
$$
which sends $[\Phi_\cov] \in H^1(Q,U_\cov)$ to the cohomology class $\sigma([\Phi_\cov]) = [\zeta]$, where
\begin{equation}\label{eq:zeta}
\zeta=d^h(\tilde \Phi\circ\theta)\in C^2(Q,C^1),
\end{equation} 
see Section \ref{Phicoc}. 
To see that the cocycle $\zeta$ belongs, in the sense of the exact sequence in Eq.~\eqref{eq:exactsequence}, to $C^2(Q,E)$ observe that 
$d^v\zeta=0$  since $d^h\Phi_\cov=0$.  
The symmetry group $G$, which is the extension of $Q$ by $N\subset E$, is a subgroup of $\tilde G$ obtained by extending $Q$ by $E$ using the cocycle $\zeta$. This is a consequence of the way $\zeta$ is defined. If $[\Phi_\cov]=0$ then the extension class $[\zeta]=\sigma([\Phi_\cov])$ of $\tilde G$ vanishes. In this case $\tilde G$ splits, which also implies that the subgroup $G$ splits. 

\end{document}